\documentclass[12pt]{article}
\usepackage{graphicx, lutypaper, multirow, chngpage}
\usepackage{caption}
\usepackage{float}
\numberwithin{equation}{section} 
\usepackage{xcolor}
\usepackage[makeroom]{cancel} 

\begin{document}

\renewcommand\tablename{{\small Table}}

\renewcommand{\vec}[1]{{\boldsymbol{#1}}} 
\renewcommand{\mat}[1]{\begin{pmatrix} #1 \end{pmatrix}}

\begin{titlepage}

\title{Primary Observables for\\
Indirect Searches at Colliders}

\author{Spencer Chang}

\address{Department of Physics and Institute for Fundamental Science\\ 
University of Oregon, Eugene, Oregon 97403}

\author{Miranda Chen, \ Da Liu, \ Markus A. Luty}

\address{Center for Quantum Mathematics and Physics (QMAP)\\
University of California, Davis, California 95616}

\begin{abstract}
We consider the complete set of observables for 
collider searches for indirect effects of new heavy physics.
They consist of $SU(3)_{\rm C}\times U(1)_{\rm EM}$ invariant 
interaction terms/operators that parameterize deviations from the Standard Model.
We show that, under very general assumptions, 
the leading deviations  from the Standard Model are given by a 
finite number of `primary' operators,  
with the remaining operators  
given by `Mandelstam descendants' whose effects
are suppressed by powers
of Mandelstam variables divided by the mass scale $M$ of the
heavy physics.
We explicitly determine all 3 and 4-point primary  operators 
relevant for Higgs signals at colliders by using the  correspondence between on-shell amplitudes
and independent operators.
We give a detailed discussion of the methods used to obtain this result, 
including a new analytical method for determining the independent 
operators.
The results are checked using the Hilbert series that counts independent 
operators.
We also give a rough sketch of the phenomenology, including unitarity
bounds on the interaction strengths and rough estimates of their 
importance for Higgs decays at the HL-LHC.
These results motivate further exploration of Higgs decays to
$Z\bar{f}f$, $W\bar{f}f'$, $\gamma \bar{f}f$, 
and $Z\gamma\gamma$. 
\end{abstract}
\end{titlepage}

\noindent
\section{Introduction}
Searches for physics beyond the standard model  (BSM) 
fall roughly into two classes:
direct searches for signals from new particle production,
and searches for indirect effects of new particles that are too
heavy to be produced.
These search strategies are
complementary, and both are important.
In this paper we will focus on experiments looking for indirect
effects of new heavy physics, with an emphasis on channels that
include the Higgs boson.
Even if the LHC does not discover new physics, the constraints
from these searches will guide further exploration of the
energy frontier, similar to the precision measurements performed
at LEP.

The main points of this paper are the following:
\begin{itemize}
\item
We propose a theoretical framework for precision measurements
that is meant to be intermediate between experimental searches
and theoretical interpretations (such as SMEFT, HEFT, or 
theoretical models).
\item
All local vertices that parameterize
deviations from the Standard Model (SM) can be written as
a linear combination of `primaries' and `descendants.'
There are finitely many primaries, and descendants are (roughly)
higher-derivative corrections to the primaries.
\item
If these vertices arise from integrating out heavy physics at
the scale $M$, the effect of the descendants is expected to be 
suppressed by powers of $E^2/M^2$ relative to the primary 
operators, where $E$ is the energy
scale of the experiment.
If the primary vertices are unsuppressed, the descendants can be
neglected, and the leading BSM effects are parameterized by
the primaries.
This leads to a larger (finite) basis of `leading' deviations 
from the SM than either HEFT or SMEFT, 
with a clear physical motivation.
\item
The classification of the primary operators was initiated in
\Refs{Durieux:2019eor, Durieux:2020gip} 
(where they are called `stripped contact terms').
This paper extends these results to operators of arbitrarily
high dimension, as well as operators with identical particles,
and operators with both massive and massless particles.
We give the complete determination of all primary
operators relevant for Higgs phenomenology.
This is in itself a nontrivial theoretical problem, and we have
developed new tools to do this.
\item
For each primary operator, we estimate the scale at which the
operator gives rise to violation of tree-level unitarity at high scales.
This is used to determine the theoretically motivated range
for the coupling constants of the primary vertices.
\item
Although this paper is primarily focused on the theoretical problem
of determining the primary operators, we give rough estimates for
the importance of primary operators for Higgs decays.
We identify a number of vertices that occur at higher orders in
the SMEFT expansion that can potentially give observable effects.
\end{itemize}

\noindent
In the remainder of this introduction, we expand on the points
above.

We are considering models where the 
leading deviations from the SM arise from particles with
mass of order $M$ that are too heavy to be directly produced in
experiments.
In this case, the effects of the heavy particles on experimental
observables with energy scale $E \ll M$ can be parameterized by
adding additional local operators to the effective Lagrangian.
In this paper, we argue that very general physical considerations
lead us to expect that the leading observed deviation from the
SM comes from so-called `primary' vertices.

The coefficient of a local interaction generated by heavy particles
depends on their mass $M$, as well as the 
couplings of the heavy particles to SM fields.
A single exchange of a heavy particle will generate an infinite
series in derivatives from expanding the propagators of the heavy
particles, for example
\[
\eql{propexpand}
\frac{1}{s - M^2} = {-\left[ \frac{1}{M^2} + \frac{s}{M^4} +
\frac{s^2}{M^4} + \cdots \right]},
\]
where $s \sim E^2$ is the energy scale of the experiment.
In this example, it is clear that the subleading terms in the
expansion will be subdominant, and it is sufficient to keep
only the first term in the expansion as long as we are searching
for the leading deviation from the SM.
This is the idea that we are attempting to make precise for
general local vertices from a `bottom-up' perspective.

To do this, we must take into account the fact that local 
operators are a redundant description of new physics effects, 
due to the freedom to perform
field redefinitions and integration by parts.
Identifying a complete and independent basis of operators is
very nontrivial.
In this paper, we use the fact that there is a one-to-one
correspondence between local operators and local on-shell 
amplitudes \cite{Elvang:2010jv,Shadmi:2018xan,Ma:2019gtx,Durieux:2019eor,
Durieux:2019siw,Durieux:2020gip,AccettulliHuber:2021uoa}.
The correspondence is simply that the on-shell amplitude is the
Feynman rule for the corresponding operator.
In this correspondence, the integration by parts
ambiguity for operators corresponds to momentum conservation
in the amplitude, 
and the freedom to perform field redefinitions in the operator basis 
is fixed by the on-shell conditions of the amplitudes.
The parameterization in terms of amplitudes is not only 
technically convenient; it has physical meaning because
the corresponding amplitudes give an estimate of the contribution
of the operator to physical processes, even if the particles
involved are not on shell.

Using the correspondence with amplitudes, we now define what we mean by
`primary' and `descendant' operators/amplitudes.
A general local on-shell amplitude can be written as a linear 
combination of `spin structures' consisting of Lorentz invariant products
of momenta and particle wavefunctions, times a 
series expansion in the Mandelstam invariants $p_i \cdot p_j$,
where $p_i$ are the momenta of the particles.
For 3-point functions, all the kinematic invariants can be written
in terms of the masses of the particles, while for 4-point functions
these can be written in terms of the Mandelstam variables $s$ and $t$
and the masses of the SM particles.
An example of a BSM correction to an on-shell 4-point amplitude is
\begin{align}
& \de\scr{M}(f_1 \bar{f}_2 \to Z_3 h_4)
= \frac{c_1^{hZ\bar{f}f}}{v} (\bar{u}_{\text{L} 2} \ggap \sla{\ep}{}_3^{\ggap *} u_{\text{L} 1}) \bigg[
1 + \al_1 \frac{s}{M^2} + \be_1 \frac{t}{M^2} 
+ O(E^4/M^4) \bigg] \nn
& \qquad{} + i  \frac{c_9^{hZ\bar{f}f}}{2v^3}\epsilon_{\mu\nu\rho\sigma}(\bar{u}_{\text{L} 2} \ga^\mu u_{\text{L} 1}) (p_1-p_2)^\nu (p_3^\rho  \ep_3^{\sigma\, *}-p_3^\sigma  \ep_3^{\rho\, *})
\bigg[ 1 + \al_9 \frac{s}{M^2} + \be_9 \frac{t}{M^2}
+ O(E^4/M^4) \bigg] + \cdots
\eql{primaryopsdefn}
\end{align}
where the numbering scheme matches the convention in the later table.  
Here, the first term in each line defines the `primary' vertex,
while the infinite series in Mandelstam variables are the
`descendants.'
Although the choice of primary operators is not unique, any choice defines a basis for the same linear space of on-shell amplitudes.

We now consider the relative importance of the various terms above
for low-energy experiments.
%
%
%
%
The couplings $c_1$ and $c_9$ depend on the couplings of the heavy 
particles to the SM fields, as well as the mass scale $M$ of the
new physics.
However, very generally, we expect that the 
coefficients $\al_{1,9}$ and $\be_{1,9}$ 
will be of order unity.
The reason is that the higher order terms in the expansion in
Mandelstam variables probes the kinematic corrections to the
amplitudes coming from the momentum expansion of heavy 
propagators, as in \Eq{propexpand}.
This tells us that we expect that we can neglect the effects 
proportional to $\al_1$ and $\be_1$ compared to the
term proportianal to $c_1$,
but this argument gives us no information about whether 
$c_1$ is expected to give larger effects than $c_9$.
The second operator has a higher mass dimension than the first,
but the coefficient $c_1$ may be suppressed
compared to $c_9$ due to the structure of the couplings of the new
physics to the SM fields.

The primary effective operators that correspond to 
\Eq{primaryopsdefn} are given by
\[
\de\scr{L} \sim \frac{c_1^{hZ\bar{f}f}}{v} h Z_\mu \bar{f}_L \ga^\mu f_L
+i \frac{c_9^{hZ\bar{f}f}}{v^3}  h  \widetilde{Z}_{\mu\nu} \big( \bar{\psi}_L\ga^\mu 
\overset\leftrightarrow{\d}{}^{\nu} \psi_L \big)
+ \cdots
\]
Note that the operators are written in terms of physical
mass eigenstate fields, as in HEFT.
Although there is no {\it a priori} limit on the dimensionality
of the primary operators, there are finitely many such operators
that are relevant for collider searches,
because such searches are only 
sensitive to 3- and 4-point functions.
Therefore, we only need to consider functions with up to 4
fields, and the expansion in derivatives is truncated if 
we drop the descendants.%
\footnote{Note that the derivative expansion is in general ill-defined
until an operator basis is specified, since the equations of
motion (field redefinitions) relate operators
with different numbers of derivatives.
We are defining the derivative expansion using the correspondence
with on-shell amplitudes.}
The result is that there is a finite basis of 
primary operators relevant for collider searches.

The argument that primary operators dominate over descendants
can fail if the fundamental physics has special features that suppress
the primary operators, but not their descendants.
This occurs in models where the Higgs boson is a pseudo
Nambu-Goldstone boson (PNGB). This important case can be treated in the 
strongly interacting light Higgs (SILH)  EFT framework \cite{Giudice:2007fh}. 
In those models, the dominant interactions of the
Higgs field are invariant under a shift symmetry
$h \to h + \la + \cdots$, and therefore
operators involving derivatives of $h$ can be more important
than operators with fewer derivatives.
For example, Ref.~\cite{Azatov:2015oxa} showed that in such models, the coupling
$\d^\rho h \d_\rho h G^{\mu\nu} G_{\mu\nu}$ is phenomenologically
relevant, and is more important
than $h^2 G^{\mu\nu} G_{\mu\nu}$, even though the former is a
descendant of the latter.
Another exception comes from the scenario discussed in \cite{Liu:2016idz},
where multipole interactions of composite gauge bosons can be enhanced
relative to monopole interactions by a factor of a strong coupling.
These examples show that we cannot claim that the primary operators
always dominate.
Even for `generic' new physics models with no special structure, 
it is possible that the primary operators could be suppressed by 
accidental cancelations.
Although the primary operators do not give the leading observables
for all possible UV models, 
they do give a large number of physically motivated observables
that we believe are worth investigating further.

We emphasize that the framework that we are proposing is an
intermediate step to connect experimental observations
with theoretical models or EFTs such as SMEFT~\cite{Buchmuller:1985jz,Grzadkowski:2010es,Brivio:2017vri,
Contino:2013kra,Elias-Miro:2013mua,Alonso:2013hga,Henning:2014wua} 
or HEFT~ \cite{Feruglio:1992wf,Alonso:2012px,Buchalla:2013rka,Brivio:2016fzo,Sun:2022ssa}.
Assuming that the dominant BSM effects come from tree-level
diagrams involving the primary vertices, they give well-defined
theoretical predictions for experimental searches 
that can be used to optimize
searches and report results.%
\footnote{The BSM vertices should be considered to be renormalized
at the scale $E$ of the experiment, and therefore include renormalization
group running effects between the scales $M$ and $E$.
For this paper, we will neglect all other loop contributions from the BSM interactions.
}
At the same time, these vertices can be computed in any theoretical
model or EFT, so the experimental results can be used to constrain
these models.
In this approach, the primary operators are treated as
independent observables in experimental searches.
Any theoretical model will have many correlations between these
observables that depend on the details of the physics at the
scale $M$.
However, given the large number of possible models of new physics,
we believe that it is sensible to take a `bottom-up'
point of view and treat the operators 
in $\scr{L}_\text{BSM}$ as independent observables for purposes
of performing individual collider searches.
Correlations between different observables in a given model can
be taken into account in global fits, while experimentalists can
focus on making the best measurement of as many observables as
possible.

We now comment on the relation this paper to previous work.
Ref.~\cite{Gupta:2014rxa} identified the leading contact interactions
arising from dimension-6 SMEFT operators, calling them `BSM primaries'
(see also the `Higgs basis'  \cite{Falkowski:2001958} and `Pseudo-Observables' frameworks \cite{Gonzalez-Alonso:2014eva, Gonzalez-Alonso:2015bha}).
Our work can be viewed as generalizing this 
to all orders in the SMEFT expansion.
There are a number of motivations to go beyond leading order in the SMEFT
expansion.
Dimension-6 SMEFT parameterizes the leading effects of decoupling physics,
namely heavy particles with mass $M$ that decouple in the limit
$M \to \infty$ with dimensionless couplings held fixed.
However, dimension-6 SMEFT does not correctly parameterize the
leading effects from non-decoupling physics, for example the
models and scenarios discussed in Refs.~\cite{Galloway:2013dma,Cohen:2020xca, Banta:2021dek,Agrawal:2019bpm}.
For non-decoupling physics, $M$ cannot be much larger than 1~TeV \cite{Chang:2019vez, Falkowski:2019tft},
but this is the scale being probed in many measurements at the
high-luminosity LHC.
For example, if there is a deviation in the $hZZ$ coupling of 
the size of the current $1\si$ constraints, the SM violates
unitarity at a scale of order $5$~TeV \cite{Abu-Ajamieh:2020yqi}.
For new physics at such low $M$, 
higher-dimension SMEFT operators could well be important.
For example, we believe it is important to independently measure
the $h\bar{t}t$ and $hh\bar{t}t$ couplings, even though these
are related by dimension-6 SMEFT.

Returning to the discussion of previous work,
Ref.~\cite{Helset:2020yio} analyzes observables to all orders in the 
SMEFT expansion, but only gives results for 3-point functions.
Refs.~\cite{Durieux:2019eor, Durieux:2020gip} analyze both 3-point and
4-point couplings using the massive spinor-helicity formalism.%
\footnote{Our `primary operators' are called `stripped contact terms' in that
work.}
They gave results only for the case that all particles are massive
and distinguishable, and did not identify primary operators to arbitrarily high
dimension for all 4-point functions.
We have tackled the additional subtleties required to treat massless and/or
indistinguishable particles, and give explicit results that we believe
are complete.
Where our results overlap, we agree with \cite{Durieux:2020gip}, 
after correcting an error in their work that was pointed out by us.

This paper is organized as follows.
In Sec.~\ref{sec:Scope} we give an overview of the three and four-point 
interactions that are needed to parametrize Higgs signals at colliders.  
In Sec.~\ref{sec:independence}, we give details about how we enumerate 
amplitudes, determine the independent 
primaries, and check the counting using Hilbert series.
In Sec.~\ref{sec:pheno}, we discuss the unitarity 
and precision electroweak constraints on the 
coefficients of these operators and give estimates for the corrections 
to Higgs decay rates.  
In Sec.~\ref{sec:Results}, we list the primary operators for three and four-point amplitudes and based on the unitarity bounds and
rough phenomenology estimates, 
determine which operators for Higgs decay are interesting at the HL-LHC.  
In Sec.~\ref{sec:conclusions} we give our conclusions and discuss future directions.

\section{Scope of Paper}
The aim of this paper is to classify the primary operators
that are relevant for Higgs signals at hadron and lepton colliders.
Specifically, we focus on all 3- and 4-point couplings that
are relevant for Higgs decays, di-Higgs production, and Higgs 
associated production.
In this section, we specify the couplings that we will study
in the remainder of the paper.
We will also define our notation and normalization conventions,
and comment on ambiguities in the operator basis associated with
off-shell 3-point couplings.

\label{sec:Scope}
\begin{figure}[!t]
\centerline{\begin{minipage}{0.8\textwidth}
\centering
\centerline{\includegraphics{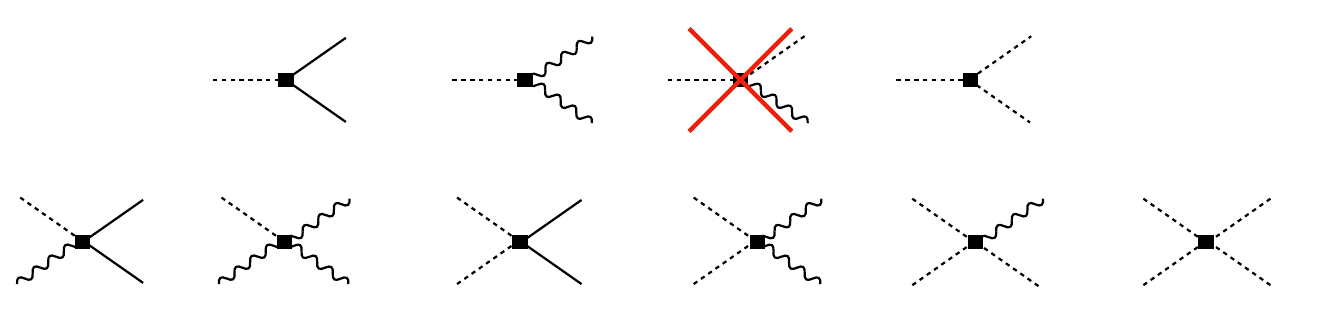}}
\caption{\small Three and 4-point couplings relevant for
Higgs decays and Higgs production.
Dashed lines denote the Higgs particle,
solid lines denote fermions, and
wavy lines denote any of the SM gauge bosons
$\ga$, $g$, $W$, or $Z$.
The crossed-out diagrams are not relevant because they vanish
on shell (see text).}
\label{fig:h34}
\end{minipage}}
\end{figure}

\subsection{Topologies and Couplings}
The 3-point and 4-point couplings that involve at least one Higgs boson
are shown in Fig.~\ref{fig:h34}.
The most general couplings compatible with
$SU(3)_\text{C}\times U(1)_\text{EM}$ gauge invariance are 
\begin{subequations}
\[
\eql{h3pt}
 \text{3-point}: & \; h\bar{f}f, hZZ, hWW, hZ\ga, h\ga\ga, hgg,
 \xcancel{\vphantom{\ga}hhZ}, \xcancel{hh\ga}, hhh, \\[5pt]
 \begin{split}
 \eql{h4pt}
 \text{4-point}: & \; hZ\bar{f}f, hW\bar{f}f', h\ga \bar{f}f, hg \bar{f}f,
 \\
 & \; hWWZ, hZZZ, hWW\ga, hZZ\ga, hZ\ga\ga, hZgg,
 h\ga\ga\ga, h\ga gg, hggg,
 \\
 & \;   hh\bar{f}f, hhWW, hhZZ, hhZ\ga, hh\ga\ga, hhgg, 
 \\
 & \; hhhZ, hhh\ga, hhhh.
 \end{split}
\]
\end{subequations}
Some of these three-point couplings vanish on-shell,
and we have crossed these out above.%
\footnote{These operators can be 
nonzero off-shell, but field redefinitions allows us to eliminate 
them in favor or
redefining the 4-point functions, as explained in \sec{onshell3} below.}

\begin{figure}[!t]
\centerline{\begin{minipage}{0.8\textwidth}
\centering
\centerline{\includegraphics{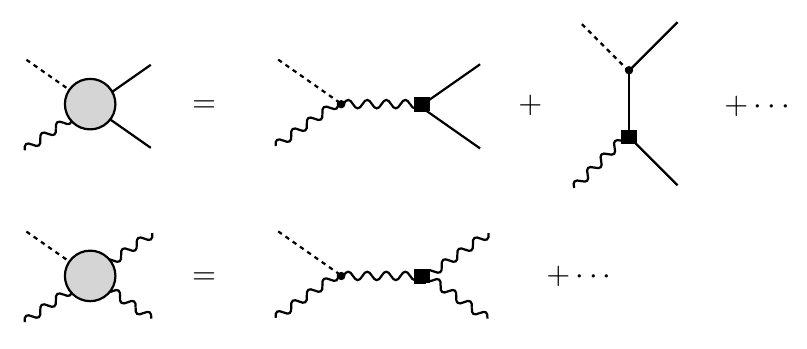}}
\caption{\small 
Exchange contributions to 4-point amplitudes
involving BSM 3-point couplings that do not
contain the Higgs.
The notation is the same as Fig.~\ref{fig:h34}.
We do not show diagrams involving Higgs exchange that involve
3-point functions already shown in Fig.~\ref{fig:h34}.
We also do not show diagrams involving two Higgses and a neutral gauge bosoon, 
since these vanish on shell (see text).}
\label{fig:noh34}
\end{minipage}}
\end{figure}

In addition, there are 3-point couplings that do not involve the
Higgs which contribute to some of the 4-point processes that get
contribution from the couplings in \Eq{h4pt}.
These are shown in Fig.~\ref{fig:noh34}.
These couplings are given by
\[
\begin{split}
\eql{noh3}
  & Z\bar{f}f, W\bar{f}f', \ga \bar{f}f, g\bar{f}f, 
  \\
  & WWZ, 
  \xcancel{\vphantom{\ga^\dagger}ZZZ},
  WW\ga, 
  \xcancel{ZZ\ga},
  \xcancel{\vphantom{\ga^\dagger}Z\ga\ga},
  \xcancel{\vphantom{\ga}Zgg}, 
  \xcancel{\vphantom{\dagger}\ga\ga\ga}, 
  \xcancel{\vphantom{\dagger}\ga gg}, 
  ggg,
\end{split}
\]
where we have again crossed out couplings that are not allowed on-shell.

These interactions parameterize the BSM contributions 
to general 2-body and 3-body decays of the the Higgs boson.
They also parameterize the BSM contributions to the production of 
a single Higgs, a pair of Higgs, and Higgs associated production
via the processes
\[
\begin{split}
& (\bar{f}f, gg, W^+W^-, ZZ) \to (h, hh, hZ, h\ga, hg)
\\
& (\bar{f}f',  ZW) \to hW,
\\
& (f g, f\ga, f Z) \to hf,
\\
& fW \to f' h.
\end{split}
\]
Note that the $hhhZ, hhh\ga$ amplitudes can be used to calculate 
exchange diagrams for $hhh$ production, 
{\it e.g.}~$\bar{f}f\to (Z^*, \ga^*) \to hhh$, 
but fully characterizing the 5-point amplitude would require us to
classify the 5-point couplings $hhh\bar{f}f$.  

Because of the large number of couplings that we are considering,
we will use a uniform notation for their couplings.
The operators contributing to a 3- or 4-point coupling 
$X = ABC$ or $ABC\!\ggap D$ 
will be denoted by $\scr{O}_i^{X}$,
where $i$ runs from 1 to the number of primary operators of type $X$.
For a primary operator $\scr{O}$ with mass dimension $d(\scr{O})$, 
we write the coupling as
\[
\eql{cO}
\De\scr{L}_\text{BSM} = \frac{c_{\scr{O}}}{v^{d(\scr{O}) - 4}} 
\scr{O},
\]
where $v = 246\GeV$ is the Higgs VEV,
and $c_\scr{O}$ is a dimensionless coefficient.
Note that if $c_\scr{O} \sim 1$ we expect the effects of the inserting
such an interaction into an electroweak process to be roughly
of order the SM contribution, since in that case all couplings
are order unity, and all mass scales are of order $100\GeV$.
For operators that are present in the SM Lagrangian, the 
coupling $c_{\scr{O}}$ is related to the associated `$\ka$ parameter' by
\[
\ka_{\scr{O}} = \frac{c_{\scr{O}} + c_{\scr{O}}^{\text{(SM)}}}{c_{\scr{O}}^{\text{(SM)}}},
\]
where $c_{\scr{O}}^{\text{(SM)}}$ is the coefficient of $\scr{O}$
in the SM Lagrangian.

\subsection{Off-Shell Ambiguities}
\scl{onshell3}
The correspondence between local on-shell amplitudes and EFT couplings
completely removes any basis ambiguity as long as the EFT couplings are
used at tree-level and on shell.
However, some of our processes of interest involve 3-point couplings
where particles are exchanged and thus potentially off-shell.
In this case, there are residual ambiguities in the basis.
These are straightforward to remove, but we discuss them here
for completeness.

To explain the point, it will be sufficient to consider a simple
example, the coupling $h\bar{f}f$.
If all particles are on shell, then this interaction is equivalent 
to the higher-derivative couplings
\[
(\Box h) \bar{f} f, \quad
h \bar{f} (i\gap \sla{\d} f), \quad
\ldots
\]
For off-shell kinematics, these operators parameterize `form factor'
corrections to the minimal on-shell coupling $h\bar{f}f$.
They parameterize the ambiguity in continuing the 
coupling $h\bar{f}f$ off-shell.
We can use field redefinitions to reduce any linear combination of such
couplings to the minimal three-point
coupling $h\bar{f}f$ \cite{Coleman:1969sm,Coleman:1985rnk,Georgi:1991ch, Arzt:1993gz}.
However, making such a field redefinition also shifts the values of some
of the 4-point couplings so that amplitudes that involve both the 3-point
and the 4-point couplings remain invariant.
The conclusion is that the choice of basis for 3-point functions is part of
the definition of the basis for the 4-point couplings.  Said in another way, if we allow for the most general local and on-shell 3 and 4-point interactions, then using them in Feynman diagrams generate the most general 3 and 4-point on-shell amplitudes
in an expansion in Mandelstam invariants.

\section{Independence of Operators/Amplitudes}
\label{sec:independence}
In this section, we explain the methods we used to determine
a basis for the 
independent primary operators.
This is done in 3 steps:
\begin{itemize}
\item
Enumerating an over-complete basis of amplitudes
\item
Determining the independent primary amplitudes
\item
Checking the result against the Hilbert series counting
\end{itemize}
We will give a short summary of each of these steps
before going into the details
in the subsections below.

The first step is to find an over-complete basis of local amplitudes
for a given process.
These basis elements are scalar monomials in the momenta and wavefunctions
of the particles involved.
They are Lorentz invariant, so the indices are contracted using the metric
and the Levi-Civita tensor.
When there are no indistinguishable particles, we can omit monomials where the momenta are contracted with other momenta,
since these can be written in terms of Mandelstam invariants and masses.
Operators with indistinguishable particles can be treated by
appropriately symmetrizing these amplitudes, as we will discuss below.
In this way, we obtain a finite number of amplitudes such that any
local amplitude is a linear combination of these amplitudes and their
Mandelstam descendants.
This step is done by hand, and in some cases we used 
{\tt Mathematica}~\cite{Mathematica} to enumerate the index contractions.

The second step is to find the independent primary interactions.
The fact that these are parameterized by on-shell amplitudes
turns this into a problem about linearly independent functions.
We proceed order by order in the number of powers of momenta.
Note that the number of momenta determines the mass dimension of the
amplitude (and the corresponding EFT operator), so we are also
working order by order in the operator dimension.
We first determine the linearly independent amplitudes of lowest
dimension that do not contain inner products of momenta.
We look for linear relations of the form
\[
\eql{ampredund}
\sum_a C_a(m) \scr{M}_a(p, s, m) = 0,
\]
where the basis amplitudes are denoted by $\scr{M}_a(p, s, m)$, where $p$ denotes
the momenta, $s$ the spins, and $m$ the masses of the particles.
The notation reminds us that the coefficients in the linear relations
can depend on the masses, but not the momenta and spins of the 
external particles.
These amplitudes  have no Mandelstam factors are thus are guaranteed to be primary amplitudes, since there is no
operator that they can be descendants of.
Then we consider operators of higher dimension, including Mandelstam
descendants of primary operators found in earlier steps that have 
the same dimension.
Eventually, we reach a dimension where all of the amplitudes
at that dimension are linear combinations of the Mandelstam descendants
of operators we already have.%
\footnote{For technical reasons, we 
have not been able to do this for operators involving
3 identical particles.
In that case, we used the Hilbert series to tell us when to stop.
The details are discussed below.}
At that point, we know that we have found all of the primary amplitudes.
We used several methods to find the linearly independent amplitudes,
including a new analytic method, and these are described below.

Finally, we compare the results to the Hilbert series counting of
operators of different dimension \cite{Lehman:2015via,Henning:2015daa,Lehman:2015coa,Henning:2015alf,Henning:2017fpj, Graf:2020yxt, Graf:2022rco,Sun:2022aag}.
The Hilbert series gives a direct counting of the primary operators
up to certain redundancies, which we review below.

We now turn to a detailed description of each of these steps.

\subsection{Enumerating the Local Amplitudes}
\scl{sec:enumerating}
The first step is to enumerate all
possible local amplitudes of a given topology and symmetry that do not
involve any Mandelstam invariants.
We will explain the procedure using the example of the $hZ\bar{f}f$
coupling.
The most general form of the corresponding amplitude is
\[
\eql{hZffexamp}
\scr{M}(f_1 \bar{f}_2 \to Z_3 h_4) 
= \bar{v}\sub{2} \Gamma^\mu u\sub{1} \ep_{3\mu}^*. 
\]
The choice of the channel is arbitrary, and does not affect the results.%
\footnote{We can also choose the masses arbitrarily, as long as we do not
take the massless limit.}
We do not use massive spinor-helicity variables because
momentum conservation is a quadratic constraint in terms of them, 
while if we work with 4-momenta we can simply write all possible
functions of the 3 independent momenta.
Also, the Mandelstam variables are manifest when the amplitude is written in
terms of the 4-momenta.

For the amplitude \Eq{hZffexamp}, the problem reduces to enumerating all
possible $\Ga^\mu$.
This is obtained by forming all possible 4-vectors formed from
$p_{1,2,3}^\mu$ and $\ga^\mu$ with indices contracted with the spacetime
metric and up to one power of the Levi-Civita tensor
(since products of Levi-Civita tensor can be written
in terms of Kronecker deltas).
We omit terms where the momenta are contracted with other momenta,
since these are Mandelstam descendants of other amplitudes.
This gives a finite list of operators that includes all primary operators.
In this way, we find
\begin{align}
\Gamma^\mu =& c_1 p_1^\mu + c_2 p_2^\mu + c_3 p_1^\mu \ga_5 + c_4 p_2^\mu \ga_5 + c_5 \ga^\mu + c_6 p_1^\mu \sla{p_3} +  c_7 p_2^\mu \sla{p_3} \nonumber \\
& + c_8 \ga^\mu \ga_5 + c_9 p_1^\mu \sla{p_3}\ga_5 +  c_{10} p_2^\mu \sla{p_3}\ga_5 + c_{11} \ga^{\mu\nu} p_{3\, \nu} \nonumber \\
& + \ep^{\mu\nu\rho\si} p_{1\,\nu} p_{2\,\rho} p_{3\, \si} \left(c_{12} + c_{13} \ga_5 + c_{14} \sla{p_3}\right)  + \ep^{\mu\nu\rho\si} \ga_\nu \left(c_{15} p_{1\, \rho} p_{2\, \si}+c_{16} p_{1\, \rho} p_{3\, \si}+c_{17} p_{2\, \rho} p_{3\, \si}\right) \nonumber \\
& + \ep^{\mu\nu\rho\si} p_{1\,\nu} p_{2\,\rho} p_{3\, \si} \left(c_{18} \sla{p_3}\ga_5 \right)  + \ep^{\mu\nu\rho\si} \ga_\nu \ga_5 \left(c_{19} p_{1\, \rho} p_{2\, \si}+c_{20} p_{1\, \rho} p_{3\, \si}+c_{21} p_{2\, \rho} p_{3\, \si}\right) \label{eqn:allhZffamps} \\
& + c_{22} \ep_{\nu\rho\si\ga} \ga^{\mu\nu} p_1^\rho p_2^\si p_3^\ga +
\ep^{\mu\nu\rho\si} \ga_{\nu\ga} p_3^\ga (c_{23} p_{1\, \rho}p_{2\, \si}+c_{24}p_{1\, \rho}p_{3\, \si}+c_{25} p_{2\, \rho}p_{3\, \si}) \nonumber \\
& +\ep^{\mu\nu\rho\si} \ga_{\nu\rho}(c_{26} p_{1\, \si}+c_{27} p_{2\, \si}+c_{28} p_{3\, \si})\nonumber \\
& + \ep^{\al\be\ga\de}\ga_\al p_{1\, \be} p_{2\, \ga} p_{3\, \de}\left(c_{29} p_1^\mu + c_{30} p_2^\mu+ c_{31} p_1^\mu \ga_5+ c_{32} p_2^\mu \ga_5\right). \nonumber
\end{align}
Note that terms containing $\ga_5$ and $\ga^\mu\ga_5$ implicitly contain one
power of the Levi-Civita tensor since 
$\ga_5 \propto \ep_{\mu\nu\rho\si} \ga^\mu \ga^\nu \ga^\rho \ga^\si$.
We have omitted terms that can obviously be simplified by equations of motion,
for example $\sla{p}{}_1 u_1 = m_1 u_1$
and $p\sub{3} \cdot \ep_3^* = 0$. 

There are several complications that are not illustrated in the present
example.
The first involves amplitudes containing massless gauge bosons, which
for us means photons and gluons.
In operator language, there are local interactions involving massless
gauge bosons that arise from expanding covariant derivatives.
However, these do not give rise to gauge invariant local amplitudes
because they are always accompanied by exchange diagrams involving the
same interaction.
For example, the $WWZ$ BSM coupling 
$\ep^{\mu\nu\rho\si} (W_\mu^+ {}\!\! \stackrel{\leftrightarrow}{D}\sub\nu \!\!{}
W^-_\nu) Z_\si$ contributes to the amplitude $WWZ\ga$ both through a
a direct 4-point coupling and an exchange diagram with a SM $WW\ga$
vertex.
In the amplitude approach, we  find the $WWZ$ local amplitude 
when characterizing the 3-point amplitudes,
and the gauge invariant operator is parameterized by the usual replacement
$\d_\mu \to D_\mu$ acting on charged fields.

The gauge invariant local on-shell amplitudes involving massless gauge bosons
must satisfy the Ward identity, and are therefore proportional to the 
combination $p_\mu \ep_\nu(p) - p_\nu \ep_\mu(p)$.
In the operator language, these correspond to gauge invariant
operators involving the field strength tensor.

Another complication that is not illustrated in our example above
occurs when we have identical particles.
For 3-point functions, this is a simple matter of symmetrizing the
amplitudes, but it is nontrivial for 4-point functions because they
can depend on Mandelstam invariants.
In this case, some of the primary amplitudes
may contain powers of the Mandelstam invariants because the
operators 
do not satisfy the appropriate Bose/Fermi symmetries without them.
For the operators we consider, we only 
have identical bosons, and we discuss the relevant cases below.

{\bf Two identical bosons:}
We want to find a basis for the primary amplitudes $\scr{M}(1234)$
where 1 and 2 are identical bosons.
We find these starting with the amplitudes where 1 and 2 are
distinguishable and then symmetrizing $1 \leftrightarrow 2$.
To do this, we first write a basis for the distinguishable amplitudes
$\hat{\scr{M}}(1234)$ that do not contain any Mandelstam invariants.
We then define the symmetric and antisymmetric combinations
\[
\hat{\scr{M}}_{\pm}(12;34) = \sfrac 12 \big[
\hat{\scr{M}}(1234) \pm \hat{\scr{M}}(2134) \big].
\]
We then construct all Mandelstam descendants of these operators
that are symmetric under $1 \leftrightarrow 2$.
This exchange acts on the Mandelstam invariants as $t \leftrightarrow u$,
so the most general such amplitude symmetric under $1 \leftrightarrow 2$
can be written as
\[
\eql{Msym12}
\scr{M}(12;34) = F(s, (t-u)^2) \gap \hat{\scr{M}}_+(12;34) 
+  (t-u) G(s, (t-u)^2) \gap \hat{\scr{M}}_-(12;34),
\]
where $F$ and $G$ are polynomial functions of their arguments.
We see that the amplitudes of the form $\hat{\scr{M}}_+(12;34)$
and $(t-u)\hat{\scr{M}}_-(12;34)$ are an over-complete basis for the
primary operators in this case,
and the higher order terms in $F$ and $G$ give the descendants.

{\bf Three identical bosons:}
Now we want to find a basis for the primary amplitudes $\scr{M}(1234)$
where 1, 2, and 3 are identical bosons.
In this case, we proceed by first symmetrizing with respect to 
$1 \leftrightarrow 2$ as above, and then symmetrize the results with respect
to the remaining symmetries.
This implies that the most general symmetric amplitude has the form
\[
\scr{M}(123;4) &= H(s, (t-u)^2) \gap \scr{M}(12;34) 
+ (2 \leftrightarrow 3) + (3 \leftrightarrow 1),
\]
where $\scr{M}(12;34)$ is a symmetrized amplitude as in 
\Eq{Msym12} and $H$ is a polynomial.
Noting that $2 \leftrightarrow 3$ implies $s \leftrightarrow t$
and $3 \leftrightarrow 1$ implies $s \leftrightarrow u$ this gives
\[
\scr{M}(123;4) &= H(s, (t-u)^2) \gap \scr{M}(12;34) 
\nn
&\qquad{}
+ H(t, (s-u)^2) \gap \scr{M}(13;24)
+ H(u, (t-s)^2) \gap \scr{M}(32;14).
\]
We can therefore start with the primary operators invariant under the
symmetry $1 \leftrightarrow 2$ and expand in powers of the Mandelstams:
\[
\begin{split}
\text{dimension $d$}\quad:\quad & \scr{M}(12;34) + \scr{M}(13;24) + \scr{M}(32;14),
\\
d+2\quad:\quad & s \scr{M}(12;34) + t \scr{M}(13;24) + u \scr{M}(32;14),
\\
d+4\quad:\quad & s^2 \scr{M}(12;34) + t^2 \scr{M}(13;24) + u^2 \scr{M}(32;14),
\\
& (t-u)^2 \scr{M}(12;34) + (s-u)^2 \scr{M}(13;24) + (t-s)^2 \scr{M}(32;14),
\\
\vdots \quad\quad\quad\ & \qquad\ \ \vdots
\end{split}
\]
The amplitudes generated in this way are not guaranteed to be Mandelstam
descendants of primary operators of 3 identical particles.
Such descendants have the form
\[
\text{descendants}\quad:\quad
\scr{M}(123;4) = J(stu, s^2 + t^2 + u^2) \gap \hat{\scr{M}}(123;4),
\]
where $\hat{\scr{M}}$ is a primary amplitude
and $J$ is a polynomial.
(Note that $s + t + u = 3m_1^2 + m_4^2$.)

Because of this issue, we cannot claim that we have rigorously enumerated all
primaries to arbitrarily high mass dimension.
The Hilbert series determines the maximum dimension of the primaries 
if we assume that there are are no relations among operators at lower
dimension (see discussion below).
The results we obtain are compatible with the Hilbert series, so
this would require a cancelation in the Hilbert series
between the new primary operators and a constraint that appears at the
same mass dimension.
This appears to be unlikely, but we cannot rigorously rule it out.
We emphasize that our methods correctly classify all the operators
up to the highest dimension that we checked.
For example, we have determined all operators of the form $h\ga\ga\ga$
and $hggg$ up to dimension 15, and we will see
that this is more than sufficient for
the phenomenology of Higgs decays at the HL-LHC.

\subsection{Independence of Amplitudes: Numerical Methods}
We now describe the methods used to determine which of the amplitudes
are independent.
This means that we have to find all 
linear redundancies of the form \Eq{ampredund}.
In this section we describe 
`brute force' numerical methods similar to those used in previous works \cite{DeAngelis:2022qco}.   

We start with a basis of amplitudes $\scr{M}_a$ with $a = 1, \ldots, n$.
The first approach is to construct an $n \times N$ matrix $X$
whose rows consist of the values of $\scr{M}_a$
for $N \gg n$ values of $p$ and $s$ and at fixed values for the masses.
This matrix can be written as
\[
X_{a\gap (p, s)} = \scr{M}_a(p, s), 
\]
where the index $(p,s)$ runs over $N$
kinematic configurations $p$,
including all possible choices of the helicities $s$
for each configuration.  
For each linear redundancy \Eq{ampredund}, this matrix satisfies 
$C\cdot X = 0$, so the redundancies are associated with the singular
values of $X$.

Equivalently, we can consider a rectangular matrix $Y$ whose
columns are given by derivatives of the amplitudes with
respect to the independent
kinematic variables, evaluated at a canonical
kinematic point $p_0$:
\[
Y_{(n,s)\gap a} \sim \frac{\d^n}{\d p^n} 
 \scr{M}_a(p, m) \Big|_{p \, = \, p_0}.
\]
Here the notation $\d^n/\d p^n$ is schematic: it means that
we consider a large number of mixed partial derivatives with
respect to the independent kinematic variables (see below).
We again include all possible choices of the spin variables $s$ for
each $\d^n/\d p^n$.
We expect that this will work for any choice of kinematic point
$p_0$, but we chose to expand the amplitudes around threshold in several
channels.

We find that both of these methods work well for moderately
large matrices, typically less than around 1000 columns.
However, for sufficiently large matrices, the numerical methods
will find more `nonzero' singular values because of the effects of
round-off errors in the numerical calculation.
This can be addressed using a smaller numerical tolerance,
and checking for robustness of the results by looking at different kinematic
configurations.

\subsection{Independence of Amplitudes: Analytical Method}
The shortcomings of the numerical approaches described above
motivated us to develop an analytical approach,
which we now describe.
To explain it, we will need to be specific about the 
kinematic variables involved.
In the center of mass frame for a $12 \to 34$ process, we can write the momenta as
\[
\eql{ps}
p_1^\mu = \mat{E_1 \\ 0 \\ 0 \\ p_i},
\quad
p_2^\mu = \mat{E_2 \\ 0 \\ 0 \\ -p_i},
\quad
p_3^\mu = \mat{E_3 \\ 0 \\ p_f \sin\th \\ p_f \cos\th},
\quad
p_4^\mu = \mat{E_4 \\ 0 \\ {-p}_f \sin\th \\ {-p}_f \cos\th},
\]
where
\begin{subequations}
\[
E_1 + E_2 &= E_3 + E_4 = E_\text{cm},
\\
|\vec{p}_1| = |\vec{p}_2| &= p_i,
\quad{}
|\vec{p}_3| = |\vec{p}_4| = p_f.
\]
\end{subequations}
and $E_{k} = \sqrt{|\vec{p}_k|^2 + m_k^2}$, $k = 1, 2, 3, 4$.
There are 2 independent kinematic
variables, which can be taken to be
$p_i$ and $\th$, for example.

For vector bosons, the polarization vectors can be taken
to have the form
\begin{subequations}
\eql{pols}
\[
\ep_{1,2}^\mu &= {e_{x\gap 1,2}} \mat{ 0 \\ 1 \\ 0 \\ 0 }
+ e_{y\gap 1,2} \mat{0 \\ 0 \\ 1 \\ 0}
+  \frac{e_{z\gap 1,2}}{m_{1,2}}
\mat{\pm p_i \\ 0 \\ 0 \\ E_{1,2}},
\\[10pt]
\ep_{3,4}^\mu &= e_{x\gap 3,4} \mat{0 \\ 1 \\ 0 \\ 0}
+ e_{y\gap 3,4} \mat{0 \\ 0 \\ \cos\th \\ {-{\sin\th}}}
+ e_{z\gap 3,4} \frac{1}{m_{3,4}}
\mat{\pm p_f \\ 0 \\ E_{3,4} \sin\th \\ 
E_{3,4} \cos\th}.
\]
\end{subequations}
Here $e_{x,y}$ are the coefficients of the transverse
polarizations (linear combinations of helicity $\pm 1$), 
while $e_z$ is the coefficient for the
longitudinal polarizations (helicity 0).
For massless vectors, only the transverse polarizations
are present.

Let us first consider a 4-point amplitude involving only vector
and scalar particles (no fermions).
From \Eqs{ps} and \eq{pols}, we see that these are polynomials in the
variables
\[
\eql{thevars}
p_i,\ p_f,\ E_{1,2,3,4},\ \sin\th,\ \cos\th.
\]
If these variables were independent of each other,
then finding the linear redundancies \Eq{ampredund}
would be a simple matter of requiring that the
coefficient of each monomial vanishes.
However, there are in fact only 2 independent variables.
Nonetheless, we show that there is a sense in 
which we can in fact treat the amplitude as a polynomial
in a set of independent variables.

To illustrate the idea, suppose that the
amplitudes were polynomials in $\cos\th$ and $\sin\th$ only.
These are not independent because of the relation 
$\cos^2\th + \sin^2\th = 1$.
We consider the polyomial to be a function of the two complex
variable $c = \cos\th$, and $s=\sin\th$.
We can use the relation to eliminate all powers of $s$ larger
than one, so that we can write the redundancy condition as
\[
\eql{redex}
0 \equiv \sum_a C_a \scr{M}_a
= P(c) + Q(c) s,
\]
where $P(c)$ and $Q(c)$ are polynomials in $c$ and since we are working with an upper bound on the operator dimension, they are also finite polynomials.
Even though $s$ and $c$ are not independent, we claim that
the constraint that the function vanishes implies that
the polynomials $P$ and $Q$ vanish identically, just as if
$s$ and $c$ were independent variables.
To see this, note that we can view the \rhs\ of \Eq{redex}
as a function of $c$ alone, with $s = \sqrt{1 -c^2}$.
For general coefficients $C_a$, there are singularities in the
complex $c$ plane that are branch cuts starting at $c = \pm 1$.
In order for this function of $c$ to vanish identically, the 
coefficient of this singularity must vanish,
which implies that the polyomial $Q$ vanishes identically:
\[
Q(c) \equiv 0.
\]
Once this condition imposed, \Eq{redex} implies
\[
P(c) \equiv 0.
\]

We can extend this method to include the full set of
kinematic variables in \Eq{thevars}.
We consider the remaining variables to be a function of $E_\text{cm}$,
which we think of as a complex variable.
Then $p_{i,f}$ are given by
\begin{subequations}
\eql{pifrational}
\[
p_i &= \frac{1}{2E_\text{cm}}
\sqrt{ \big[ E_\text{cm}^2 - (m_1 + m_2)^2 \big]
\big[ E_\text{cm}^2 - (m_1 - m_2)^2 \big]} \,,
\\
p_f &= \frac{1}{2E_\text{cm}}
\sqrt{ \big[ E_\text{cm}^2 - (m_3 + m_4)^2 \big]
\big[ E_\text{cm}^2 - (m_3 - m_4)^2 \big]} \,.
\]
\end{subequations}
These have branch point singularities at 4 points, 
$E_{\text{cm}} = \pm (m_1 \pm m_2), 
\pm (m_3 \pm m_4)$.
The energies $E_k$ can be written in terms of $E_\text{cm}$
using
\begin{subequations}
\eql{Erational}
\[
E_1 &= \frac{m_1^2 - m_2^2 + E_\text{cm}^2}{2E_\text{cm}},
\qquad
E_2 = \frac{m_2^2 - m_1^2 + E_\text{cm}^2}{2E_\text{cm}},
\\
E_3 &= \frac{m_3^2 - m_4^2 + E_\text{cm}^2}{2E_\text{cm}},
\qquad
E_4 = \frac{m_4^2 - m_3^2 + E_\text{cm}^2}{2E_\text{cm}}.
\]
\end{subequations}
We can use \Eqs{pifrational} and \eq{Erational}
to eliminate the dependence on $E_{k}$ and even powers of $p_{i,f}$.
The resulting function of $E_{\text{cm}}$ has $1/E_\text{cm}^n$
singularities, which we eliminate by multiplying by
$E_{\text{cm}}^N$ for some sufficiently large $N$.
The result has the form of a polynomial in $E_\text{cm}$,
$s$, $c$, $p_i$, $p_f$ with at most linear powers of 
$s$, $p_i$, and $p_f$:
\[
0 &= E_\text{cm}^N \sum_a C_a \scr{M}_a
\nn
&= P + Q s +  R p_i + S p_f
+ T s \gap p_i  +  U s \gap p_f 
+ V \gap p_i p_f
+  W\gap s \gap p_i p_f,
\eql{polyful}
\]
where $P, \ldots, W$ 
are polynomials in $E_\text{cm}$ and $x$.
Because $s$, $p_i$, and $p_f$ all have different singularity
structure when written as functions of $E_\text{cm}$ and $c$,
we can treat all of the variables in \Eq{polyful}
as independent when solving the constraints, which again requires that all of the polynomials separately vanish.

Extending these ideas to amplitudes involving fermions
is nontrivial because the spinor wavefunctions contain
factors of $\sqrt{E \pm p_{i,f}}$.
We were able to extend the method to amplitudes involving
2 fermions, for special choices of the fermion masses.
Taking the fermions to be the incoming particles, the
spinor wavefunctions are functions of
$\sqrt{E_{1,2} \pm p_i}$, for example
\[
(u_1)_s &= \mat{ \mat{\sqrt{E_1 - p_i} & 0 \\ 0 & \sqrt{E_1 + p_i}}
\xi_s \\[20pt]
\mat{\sqrt{E_1 + p_i} & 0 \\ 0 & \sqrt{E_1 - p_i}} \xi_s}.
\]
where $s = 1,2$ is the spin label and $\xi_{1,2}$ are a basis for
2-component spinors.
The analytic method can be extended for the following
special cases:
\begin{itemize} 
\item $m_1 = m_2$:
In this case $E_1 = E_2 = E_i$.
The amplitude is proportional to the product of
spinor wavefunctions for particles 1 and 2,
which contain the following square root structures:
\[
\big( \sqrt{E_i \pm p_i} \big)^2 &= E_i \pm p_i,
\\
\sqrt{E_i + p_i} \sqrt{E_i - p_i} &= m_1 = m_2.
\]
The constraints therefore have the same form as \Eq{polyful}.
\item $m_2 = 0$:
In this case we have 
\[
\eql{pibranch}
p_i = E_2 = \frac{E_\text{cm}^2 - m_1^2}{2 E_\text{cm}},
\]
so $p_i$ no longer has a branch cut singularity as a function
of $E_\text{cm}$, but $p_f$ does.
The spinor wavefunctions contain the following square root
structures:
\begin{subequations}
\[
\eql{spinsqrt1}
\sqrt{E_1 + p_i} &= \sqrt{E_\text{cm}},
&
\sqrt{E_1 - p_i} &= \frac{m_1 \sqrt{E_\text{cm}}}{E_\text{cm}},
\\
\eql{spinsqrt2}
\sqrt{E_2 + p_i} &= \sqrt{2p_i},
&
\sqrt{E_2 - p_i} &= 0.
\]
\end{subequations}
The amplitudes are proportional to one factor from
\Eq{spinsqrt1} and one factor from \Eq{spinsqrt2},
so the nonzero amplitudes are all proportional to
$\sqrt{E_\text{cm}} \sqrt{p_i}$.
By multiplying by $\sqrt{E_\text{cm}} \sqrt{p_i} E_\text{cm}^N$
for some $N$, the constraints can 
therefore be written as a polynomial in $E_\text{cm}$, $s$, $c$, $p_f$
that is linear in $s$ and $p_f$:
\[
0 &= \sqrt{E_\text{cm}} \sqrt{p_i} E_\text{cm}^N \sum_a C_a \scr{M}_a
\nn
&= P
+ Q s +  R p_f
+ S p_f s ,
\eql{polyfulm0}
\]
where $P, \ldots, S$ are again polynomials in $E_\text{cm}$
and $c$.
The same argument above therefore shows that we can treat
all of the variables in \Eq{polyfulm0} as independent
when solving the constraints.
\end{itemize}

\noindent
We find that both methods find the same sets of independent
amplitudes with 2 fermions, and that these methods also agree
with the numerical methods for generic masses.
This is reassuring, since we do not expect the independent
amplitudes to be different for special choices of the 
fermion masses.

To summarize, given that the redundancies require the polynomials to 
individually vanish, we can analyze the number of independent amplitudes 
by choosing the kinematic variables $E_{cm}, p_i, p_f, c, s$ 
where we treat them independently, as long as we've replaced factors 
of $s^2, p_i^2, p_f^2$ in terms of $c$ and $E_{cm}$.    
It would be interesting to generalize this analytic
argument to general amplitudes, for example involving 4 fermions.
The method relies on the fact that the singularities of the
amplitudes are simple square root branch cuts.
In comparison to  the spinor helicity formalism, the local amplitudes are
polynomials in spinor-helicity variables.
These variables are also not independent, but the constraints they satisfy
are quadratic polynomial equations.
It is natural to speculate that
this underlying structure allows us to
generalize the results above beyond special kinematic points.

\subsection{Hilbert Series}
\scl{HilbertSeries}
An important check of our results is the Hilbert series that counts the number
of independent EFT operators, described in
\Refs{Lehman:2015via,Henning:2015daa,Lehman:2015coa,Henning:2015alf,Graf:2022rco}.
The Hilbert series counts the number of operators at a given mass dimension,
taking into account symmetry constraints as well redundancies due to integration
by parts are field redefinitions.%
\footnote{We thank X. Lu for patiently explaining Hilbert
series to us.}

The Hilbert series for our trilinear interactions are the following:  
\[
\label{eqn:Hilbert3pt} 
\begin{split}
& H_{h\bar{f}f} = 2q^4,\quad
H_{h\ga Z}=H_{h\ga\ga} = H_{hgg} = 2q^5,\quad
H_{hZZ} = H_{hWW} = q^3+2q^5,  
\\
& H_{hhZ} = H_{hh\ga} =0,\quad 
H_{hhh} = q^3,
\\
& H_{\ga \bar{f}f} = 2q^5,\quad 
H_{Z \bar{f}f}=H_{W \bar{f}f'} = 2q^4 + 2q^5,  
\\
& H_{WWZ} = 5q^4+2q^6,\quad 
H_{WW\ga} = 2q^4+2q^6,\quad 
H_{ggg}=2q^6,
\\
& H_{ZZZ} = H_{ZZ\ga}=H_{Z\ga\ga}= H_{Zgg}=0.
\end{split}
\]
Here $q$ is a parameter that counts the mass dimension of the operators.  
The power of $q$ in each term is the mass dimension of the operator, and the 
coefficient gives the number of operators at that dimension.
So for example, $H_{h\bar{f}f} = 2 q^4$ implies that there are 2 operators
with dimension 4.

The Hilbert series for our four-point interactions are the following:
\[
\begin{split}
& H_{hZ\bar{f}f} = H_{hW\bar{f'}f} = \frac{2q^5+6q^6+4q^7}{(1-q^2)^2}, \quad H_{h\ga\bar{f}f} = H_{hg\bar{f}f}= \frac{2q^6+4q^7+2q^8}{(1-q^2)^2},  \\
& H_{hZ\ga\ga} = H_{hZgg}= \frac{3q^7+7q^9+2q^{11}}{(1-q^2)(1-q^4)}, \quad H_{hggg} =   \frac{2q^{7}+2q^{9}+4q^{11}+6q^{13}+2q^{15}}{(1-q^4)(1-q^6)},  \\
& H_{h\ga g g}= \frac{4q^{9}+4q^{11}}{(1-q^2)(1-q^4)} , \quad H_{h\ga\ga\ga}= \frac{2q^{11}+4q^{13}+2q^{15}}{(1-q^4)(1-q^6)},  \\
& H_{hWW\ga} = \frac{2q^{5}+14q^{7}+2q^9}{(1-q^2)^2}, \quad  H_{hZZ\ga} = \frac{8q^{7}+8q^9+2q^{11}}{(1-q^2)(1-q^4)}, \\
& H_{hWWZ} = \frac{9q^{5}+18q^{7}}{(1-q^2)^2}, \quad  H_{hZZZ} = \frac{q^5+6q^7+8q^9+7q^{11}+5q^{13}}{(1-q^4)(1-q^6)}, \\
& H_{hh\bar{f}f} = \frac{2q^5+2q^8}{(1-q^2)(1-q^4)},
\\
& H_{hhWW} = \frac{q^4+3q^6+5q^8}{(1-q^2)(1-q^4)}, \quad H_{hhZZ}  = \frac{q^4+3q^6+2q^8}{(1-q^2)(1-q^4)}  \\
&H_{hhZ\ga} =\frac{2q^6+4q^8}{(1-q^2)(1-q^4)}, \quad H_{hh\ga\ga}= H_{hhgg} = \frac{2q^6+q^8}{(1-q^2)(1-q^4)},  \\
& H_{hhhZ}= \frac{q^{7}+q^9+q^{13}}{(1-q^4)(1-q^6)} , \quad H_{hhh\ga}= \frac{2q^{13}}{(1-q^4)(1-q^6)}, 
\\
& H_{hhhh}= \frac{1}{(1-q^4)(1-q^6)}.
\end{split}
\label{eqn:Hilbert4pt}
\] 
The denominators represent the infinite series of Mandelstam descendants.
For the couplings where all particles are distinguishable, this factor is given by
\[
\frac{1}{(1-q^2)^2} = (
1 + q^2 + q^4 + \cdots )^2,
\]
which counts the series of products of the two independent Mandelstam
variables ($s$ and $t$ say).
For couplings containing indistinguishable particles, the denominator
factor is modified because the series of Mandelstam variables is
constrained by symmetry.
For example, for $hh\bar{f}f$ in the channel $hh \to \bar{f}f$,
the independent symmetric Mandelstam invariants are $s$ and $(t-u)^2$.
The denominator factor is given by
\[
\frac{1}{(1-q^2)(1-q^4)}
= (1 + q^2 + q^4 + \cdots)(1 + q^4 + q^8 + \cdots)
\]
which counts the series of products of $s$ and $(t-u)^2$.
For $hZZZ$, the independent symmetric Mandelstam invariants are
$s^2+t^2+u^2$ and $stu$, which is matched by the denominator factor
\[
\frac{1}{(1-q^4)(1-q^6)} = (1 + q^4 + q^8 + \cdots)
(1 + q^6 + q^{12} + \cdots).
\]

This suggests that the the numerator factors simply count the number of
primary operators at each dimension.
While this is the simplest interpretation, it is not necessarily correct.
The reason is that there can be relations between Mandelstam descendants
of independent primary operators.
For example, two lower dimensional primaries may become redundant at 
higher mass dimension when one includes enough Mandelstam factors.
If there are $n$ such relations that arise at dimension $d$, this is parameterized
in the Hilbert series by an infinite series
\[
\frac{-n q^d}{(1-q^2)^2}
= -n q^d (1 + q^2 + q^4 + \cdots)^2,
\]
which subtracts off the redundant terms in the Hilbert series.
(The remaining positive terms in the Hilbert series must of course ensure that 
the coefficient of each power of $q$ is positive.)
In fact, negative terms in the numerator of the Hilbert series 
appear for 4-fermion couplings, which are not considered in this work.
Although all of the coefficients in the numerators of the Hilbert series
above are positive, it is possible that there are relations at the same
mass dimension that we have new primaries.
In other words, the coefficient of $q^d$ in the numerator
is equal to the number of independent
primaries minus the number of relations between Mandeltam descendants that
appear at dimension $d$.
For all operators other than those that contain 3 identical particles,
our methods determine all primary operators up to arbitrary mass dimension
independently of the Hilbert series.
In these cases, the Hilbert series is used only as a check, and we find
that the coefficients in the numerators do in fact count the number of 
primary operators in all cases.

For the case of 3 indistinguishable particles, as discussed in 
\sec{sec:enumerating}, our methods do not guarantee that there
are no additional primary operators at dimensions higher than we have
explicitly checked.
In these cases, the primary operators we find are equal to the
coefficients of the numerators of the Hilbert series above, so 
we again have agreement with the Hilbert series.
However, we cannot exclude the possibility that at higher dimensions
there are additional primary operators with  an equal number of
additional constraints at that dimension.
Even if this is the case, we have determined all primary operators up
to dimension 13, and any additional operators are unlikely
to be phenomenologically relevant.

\section{Phenomenology from the Bottom Up}
\label{sec:pheno}
In this section, we discuss some of the basic phenomenology
of the operators that we have found.
We first show that unitarity bounds can give us an upper bound on the couplings of the SM deviations.
As emphasized in \cite{Chang:2019vez},
any new interaction that is not included in the SM
implies that tree-level unitarity is violated at some
energy scale, and this scale can be estimated without a
complete EFT framework.
Assuming an energy scale where unitarity is valid to,  enables us to to give an upper bound on couplings of the interactions.
In this section, we will  describe the assumptions and methods
that we use to obtain these bounds. 
We also give rough estimates of the size
of physical effects of the new interactions for Higgs decays.
Comparing these to the unitarity bounds gives an idea of
which operators may be plausibly large enough 
to be observed in upcoming Higgs searches.

\subsection{Perturbative Unitarity Bounds}
\scl{Unitarity}
We now describe how we place bounds on the coefficients
of the primary operators from unitarity considerations.
It is a classic result that the 
SM is the unique theory with the observed particle content
that does not violate tree-level unitarity at high energies
 \cite{Cornwall:1974km} 
 (see \cite{Liu:2022alx} for a purely on-shell derivation).
Therefore, any deviation from the SM will lead to a violation
of tree-level unitarity at some scale, which can be used to bound the
scale of new physics.
We now turn this around to determine the allowed coefficients of the
primary interactions such that the scale of unitarity violation is 
larger than some value, for example 1 TeV.
This gives a theoretical upper bound on the deviations from the SM that 
can be used to decide which searches are sufficiently motivated to
carry out.

As emphasized in \cite{Chang:2019vez,Abu-Ajamieh:2020yqi,Abu-Ajamieh:2021egq,Abu-Ajamieh:2022ppp},
the unitarity bounds can be obtained from a purely bottom-up
perspective (without assuming any EFT power counting),
but the unitarity bounds do depend on what assumptions we make
about other couplings.
To illustrate this, we consider the coupling $hh\bar{t} t$.  
We want to know whether this coupling could possibly be the
first observed sign of new physics.
In order for this to be the case, we must
assume that the BSM contribution to the $h\bar{t}t$ coupling is suppressed,
due to the greater sensitivity of experiments to this coupling.
This assumption affects the unitarity bounds on the $hh\bar{t}t$ couplings,
as we will now explain.

The strongest constraint on the $hh\bar{t}t$ coupling
from unitarity violation at the highest energies comes not from
the 4-particle amplitudes such as $hh \to \bar{t}t$, but from
higher-point amplitudes involving longitudinally polarized $W$ and $Z$ bosons.
This arises because the $hh\bar{t}t$ coupling ruins cancelations
that otherwise ensure tree-level unitarity of these higher-point 
amplitudes.
As shown in \Refs{Chang:2019vez,Falkowski:2019tft,Abu-Ajamieh:2020yqi}, 
these can be understood
at the level of the Lagrangian using the Goldstone
boson equivalence theorem~\cite{Cornwall:1974km,Vayonakis:1976vz,Chanowitz:1985hj}. 
The point is that gauge invariance implies that couplings like 
$hh\bar{t}t$  have associated dependence on the triplet of
eaten Nambu-Goldstone fields $\vec{G}$, and the amplitudes for 
the Nambu-Goldstone bosons are the same as the longitudinal 
$W$ and $Z$ bosons in the high-energy limit, which can be determined by replacing
\[
h &\to \sqrt{(v+h)^2+{\vec{G}}^2}-v,\\
\bar{t}t & \to \frac{1}{\sqrt{(v+h)^2+{\vec{G}}^2}}\left((v+h)\bar{t}t+G^0 \bar{t}i\ga_5 t -\sqrt{2} G^+ \bar{b}_R t_L -\sqrt{2} G^- \bar{b}_L t_R\right).
\]
For simplicity, we only consider amplitudes of the form $\bar{t}tG^m$ from 
$h^n \bar{t}t$ couplings, so we can then expand the expressions above to give 
\[
&\sum_n c_{t,n} h^n \bar{t} t
\nn
&\quad{}
\to \left(\frac{c_{t,1}}{2v}{\vec{G}}^2 -\frac{3c_{t,1}-2c_{t,2}}{8v^3}{\vec{G}}^4 +\frac{5c_{t,1} - 4 c_{t,2} + 2 c_{3,t}}{16v^5}{\vec{G}}^6 
+\cdots  \right) \left(\bar{t}t + \frac{G^0}{v}\bar{t}i\ga_5t\right)+\cdots 
\]
Note that $c_{t,2}$ gives rise to amplitudes of the form $\bar{t}t G^n$
for $n \ge 4$, but these can be canceled by other couplings.
Because we are assuming that $c_{t,1}$ is small, its contribution cannot
cancel the contribution to the $\bar{t}t G^4$ and $\bar{t}t G^5$
couplings, but the higher couplings can be canceled by the 
unconstrained couplings $c_{t,n}$ for $n \ge 3$.
We can therefore use the $\bar{t}t G^4$ and $\bar{t}t G^5$ couplings
to obtain a unitarity bound on $c_{t,2}$.
We see that with the assumptions that we are making, the $hh\bar{t}t$
coupling effectively behaves like a dimension 8 operator at high
energies.
This can also be understood from the perspective of SMEFT, as we
will discuss below.

In general, we compute the unitarity bounds for 4-point couplings
under the assumption that the 3-point couplings are sufficiently
small that their contribution to the unitarity bound can be
neglected.
If a deviation from the SM is observed in any channel, one would obviously
want to perform a complete analysis including all experimental 
constraints, but we believe that the bound we are
presenting is appropriate for the purpose at hand.

To calculate the unitarity bounds from higher-point processes such as
$\bar{t}t \to G^5$, we use the results of 
\Refs{Chang:2019vez,Abu-Ajamieh:2020yqi}.
We will use a simplified version of these estimates that neglects
some numerical factors of order 1.
A coupling of $n$ distinguishable scalars can be written
\[
\scr{L}_\text{int} = \frac{C_n}{v^{n-4}} \phi_1 \cdots \phi_n,
\]
and the associated scattering amplitudes are
\[
\scr{M}(\phi_1 \cdots \phi_k \to \phi_{k+1} \cdots \phi_n)
\sim \frac{C_n}{v^{n-4}}.
\]
The unitarity bound on this amplitude is \cite{Abu-Ajamieh:2020yqi}
\[
\eql{unitbound}
\scr{M}(\phi_1 \cdots \phi_k \to \phi_{k+1} \cdots \phi_n)
\lsim \frac{1}{\sqrt{\Phi_k(E) \Phi_{n-k}(E)}},
\]
where
\[
\Phi_k(E) \sim \frac{1}{8\pi} 
\left( \frac{E}{4\pi} \right)^{2k-4}
\]
is the total massless phase space for $k$ distinguishable massless 
particles with total energy $E$, where we have neglected 
a combinatoric factor $1/(k-1)!(k-2)!\,$.
By ignoring those combinatorial factors, the combination 
$\Phi_k \Phi_{n-k}$ that appears in \Eq{unitbound}
is independent of $k$, and we do not have to optimize the number of
incoming and outgoing particles.
If we require that unitarity is satisfied up to some maximum energy
$E_\text{max}$, we obtain the unitarity bound
\[
C_n \lsim 8\pi 
\left( \frac{4\pi v}{E_\text{max}} \right)^{4-n}.
\]
For a fermion coupling
\[
\De\scr{L} = \frac{C_{t,n}}{v^{n-1}} \bar{t} t \phi_1 \cdots \phi_n
\]
we have 
\[
\scr{M}(\underbrace{\bar{t} t \phi_1 \cdots \phi_k}_{k+2}
\to \underbrace{\phi_{k+1} \cdots \phi_n}_{n-k}) \sim \frac{C_{t,n}}{v^{n-1}} E,
\]
and we obtain the bound
\[
C_{t,n} \lsim 
\left(\frac{4\pi v}{E_\text{max}} \right)^{n-1}.
\]
In this way, we obtain the approximate unitarity bounds
\begin{subequations}
\eql{tthhunitaritybounds}
\[
\bar{t} t \to G^2 \ \ &:\ \  
c_{t,2} \lsim 6 / E_\text{TeV}, \\ 
\bar{t} t \to G^3 \ \ &:\ \  
c_{t,2} \lsim 20 / E_\text{TeV}^2 \\ 
\bar{t} \gap t \to G^4 \ \ &:\ \  
c_{t,2} \lsim 60 / E_\text{TeV}^3 \\
\bar{t} \gap t  \to G^5 \ \ &:\ \  
c_{t,2} \lsim 200 / E_\text{TeV}^4, 
\]
\end{subequations}
where $E_\text{TeV}$ is $E_\text{max}$ measured in TeV. 
Even though these estimates were obtained by ignoring combinatoric 
factors in the phase space and matrix elements, they agree well with the results of \cite{Abu-Ajamieh:2020yqi}, where
all such factors are included.

Which of the unitarity bounds in \Eq{tthhunitaritybounds}
is the strongest depends on the scale $E_\text{max}$.
For asymptotically large values of $E_\text{max}$, the process with the
most particles gives the strongest bound, but for low values of $E_\text{max}$
the process with the smallest number of particles dominates.
If we neglect combinatoric factors, these bounds cross at the
NDA scale $E_\text{max} \sim 4\pi v \sim 3\TeV$.
In the tables, we will give the unitarity bounds in terms of 
$E_\text{TeV}$, 
since 1 TeV is roughly the scale that has been probed by measurements at the LHC.

Although every Higgs interaction can be understood from the bottom-up 
approach described above, 
we find it convenient 
to use SMEFT operators as a proxy for 
calculating the unitarity bounds in our tables.
Specifically, for 3-point functions, we use the lowest-dimension SMEFT operator
as a proxy, while for 4-point functions, we use a combination of
SMEFT operators of lowest dimension that does not modify the 3-point functions.
This is motivated by the fact that 3-point functions are generally more
constrained by experiments.
In the example of $hh\bar{t}t$, we use a combination of the 
$H^\dagger H \bar{Q}_L \widetilde{H} t_R$ and
$(H^\dagger H)^2 \bar{Q}_L \widetilde{H} t_R$ SMEFT operators,
and assume that the
deviation in $h\bar{t}t$ is suppressed by a cancelation between them.
This could be viewed as an accidental cancelation, or it may be that
the SMEFT power counting simply does not hold for new physics at low scales.
The SMEFT approach predicts that interaction behaves as a dimension-8 operator
with at most 7-particle interactions, just
as we found from the bottom-up point of view.
When we estimate the unitarity bounds for couplings such as
$hZ\bar{t}t$ and $hW\bar{t}b$, we will assume that they come in combinations
that preserve custodial symmetry, since this gives weaker unitarity constraints. 
The fact that custodial symmetry is straightforward to incorporate in
SMEFT is another reason we make use of it.

Let us illustrate the use of SMEFT operators to obtain the unitarity
bounds with the example of the coupling $h Z_\mu \bar{t}_L \ga^\mu t_R$.
We assume that the 3-point coupling $Z_\mu \bar{t}_L \ga^\mu t_R$ is not
modified, so this requires a cancelation between the SMEFT operators
$(H^\dagger \!\gap \overset{\leftrightarrow}{D}_\mu \!\gap H) 
\bar{Q}_L \ga^\mu Q_L$
and 
$H^\dagger H 
(H^\dagger \!\gap \overset{\leftrightarrow}{D}_\mu \!\gap H) 
\bar{Q}_L \ga^\mu Q_L$.
We have
\[
\frac{c^{hZtt}_1}{v} h Z_\mu \bar{t}_L \ga^\mu t_R
\subset \frac{c^{hZtt}_1}{m_Z v^3} 
H^\dagger H (H^\dagger \!\gap \overset{\leftrightarrow}{D}_\mu \!\gap H)
\bar{Q}_L \ga^\mu Q_L,
\]
where the additional factors of $m_Z$ and $v$ 
on the \rhs\ come from expanding the Higgs doublets
and covariant derivatives.%
\footnote{From the bottom-up point of view, we can understand the
factor of $1/m_Z$ from the equivalence theorem
$Z_\mu \to \d_\mu G^0/m_Z$ at high energies.}
(We are ignoring order-1 numeric factors, since we are
performing a rough calculation.)
We see that at high energies, the unitarity growth is that of a dimension-8
operator, and that we can consider amplitudes with a maximum of 7 particles.
The fastest energy growth at high energies can be read off from the
amplitude
\[
\scr{M}(\bar{t}t \to G^4) \sim \frac{c^{hZtt}_1}{m_Z v^3} E_\text{max}^2
\lsim \bigg[ 8\pi \frac{(8\pi)^5}{E_\text{max}^4} \bigg]^{1/2}
\qquad\Rightarrow\qquad
c^{hZtt}_1 \lsim \frac{5}{E_\text{TeV}^4}.
\]
Processes with 7 particles such as $\bar{t}t \to ZG^4$ trade
one derivative (power of energy) with an additional $Z$ boson
and give a slightly weaker bound at high energies.
At lower energies, the bound comes from the processes such as
\[
\scr{M}(\bar{t}t \to G^2)
\sim \frac{c^{hZtt}_1}{m_Z v} E^2_\text{max}
\lsim 8\pi
\qquad\Rightarrow\qquad
c^{hZtt}_1 \lsim \frac{0.6}{E_\text{TeV}^2}.
\]
As mentioned above, with these approximations all of the unitarity
bounds become degenerate at $E_\text{max} \sim 4\pi v\sim 3\TeV$,
so it is sufficient to compute the one with the bounds for the processes
with the largest and smallest number of particles.

\subsection{Precision Electroweak Constraints}
Precision electroweak measurements also give stringent constraints
on corrections to the SM.
In our approach, primary operators that are not directly
constrained by precision electroweak measurements are simply
treated as independent.
For example, $\mu$ decays constrain one linear combination of
the $W\bar{\ell}\nu$ couplings, but allow large deviations in
individual couplings if there is a cancelation in the combination
that controls the $\mu$ decay rate.
From a bottom-up perspective, precision electroweak constraints
are similar to naturalness constraints, since they can be
satisfied by fine-tuning different contributions to the same
process.

However, the degree of cancelation required
to obtain an observable signal is an important factor
in deciding which observables are sufficiently well-motivated to
merit further investigation.
We therefore performed estimates of loop-induced precision
electroweak corrections, even though we are not working in a complete
EFT framework.
That is, we treat the primary operators as interaction terms in 
an $SU(3)_\text{C} \times U(1)_\text{EM}$ invariant EFT, and
estimate the size of loop corrections with a UV cutoff $\La$ that
we identify with the scale of new physics.
We have not analyzed all of the
primary operators, but we generally find that requiring the absence of 
cancelations in precision electroweak observables gives weaker constraints
than the unitarity constraints as long as we assume that the 
new physics satisfies custodial symmetry.

As an example of a strong constraint in the absence
of custodial symmetry, we consider the operator $hhZ^\mu Z_\mu$.
Closing the Higgs loop gives a quadratically divergent contribution
to the $Z$ mass. 
If this is not canceled by a custodial preserving contribution to
the $W$ mass, we obtain the constraint on the coefficient
\[
c^{hhZZ}_1 \lsim \frac{10^{-3}}{\La_\text{TeV}^2},
\]
where $\La_\text{TeV}$ is the cutoff in TeV units
and we are using the operator numbering in Table~\ref{tab:hhVV}.
If we identify $\La$ with the unitarity violating scale $E_\text{max}$,
the precision electroweak constraint is stronger than the unitarity constraint for 
$\La \lsim 40$~TeV (see Table~\ref{tab:hhVV}).
Approximate custodial symmetry can significantly weaken this constraint, but
its implementation in EFT is subtle (see \cite{Kribs:2020jgn}).
Therefore, we will not attempt to estimate corrections to precision
electroweak observables that are sensitive to custodial symmetry
violation.

We now some examples of the precision electroweak constraints
for some of the operators that are the most promising for Higgs
decay phenomenology (see \sec{PrimaryDecays} below).
For example, the CP-even operators
$hZ^\mu Z_\mu$ and $h Z^{\mu\nu} Z_{\mu\nu}$
give a 1-loop contribution to the $Z$ kinetic term, generating
a correction equivalent to the $S$ parameter.
This gives the constraints
\[
c_1^{hZZ} \lsim 20 \ggap \La_\text{TeV}^2,
\qquad
c_2^{hZZ} \lsim \frac{0.5}{1 + 0.4 \cdot \log \La_\text{TeV}}.
\]
where we have used the operator numbering in Table~\ref{tab:h3p}.
These are weaker than the corresponding unitarity constraints.

Next, we consider the CP-even $hZff$ couplings in Table~\ref{tab:hZff}.
At one loop these induce a correction to  $Zff$ couplings, which
are highly constrained by LEP.
Operators 1 and 2 induce a correction to the vector and axial-vector $Z$ couplings, and
give
\[
c_{1,2}^{hZff} \lsim \frac{0.5}{1 + 0.4 \cdot \log\La_\text{TeV}},
\]
which are comparable to the unitarity bounds for $\La \sim \TeV$,
but are otherwise weaker.
Operator 5 corrects the coupling 
$i Z^\mu \bar\psi \overset\leftrightarrow{\d}{}^\mu \psi$,
which flips the fermion helicity.
This has a weaker constraint at LEP because it does not interfere
constructively with the SM $Z$ coupling.
Using the results of \cite{Escribano:1993xr}, we find the weak constraint
\[
c_5^{hZff} \lsim \frac{60}{1 + 0.4 \cdot \log\La_\text{TeV}}.
\]
Operator 7 corrects the coupling $\d^\mu Z_\mu \bar\psi \psi$, which vanishes
on shell.
To get a nonzero correction, we must go to higher loop, and this will
give weak constraints.
Operators 9 and 11 correct the coupling 
$i\tilde{Z}_{\mu\nu} \bar\psi \ga^\mu \overset\leftrightarrow{\d}{}^\nu \psi$,
which  gives the constraint
\[
c_{9,11}^{hZff} \lsim \frac{3}{1 + 0.4 \cdot \log\La_\text{TeV}},
\]
which is weaker than the unitarity bound.

The general pattern that we find
is that the unitarity bounds are more sensitive to
the UV scale than the precision electroweak observables,
at least if we neglect the corrections to the $W$ and $Z$ masses
that violate custodial symmetry.
It would be interesting to give a more complete analysis, including constraints on CP-odd operators, but we
leave this for further work.

\subsection{Estimates for Higgs Decays}
We now perform some crude estimates determine what ranges of BSM
couplings can be probed in Higgs decays at the LHC.
Specifically, we will estimate the corrections to the branching
ratios of Higgs decays to determine which operators can give an
observable number of Higgs decays.
These couplings can then be compared to the unitarity bounds discussed
above to determine whether it is motivated to search for a particular
coupling.

We will focus on operators that are not present in the SM.
In the case where the BSM operator $\scr{O}$ modifies a Higgs
coupling, the phenomenology can be studied in the 
so-called `$\ka$ framework' \cite{LHCHiggsCrossSectionWorkingGroup:2012nn}.
The $\ka$ parameter associated to $\scr{O}$ is given by
\[
\ka_{\scr{O}} = \frac{c_{\scr{O}}^{\text{(SM)}} 
+ c_{\scr{O}}}{c_{\scr{O}}^{\text{(SM)}}}.
\]
Projections for the sensitivity of the HL-LHC to various $\ka$ parameters
can be found in Ref.~\cite{Cepeda:2019klc}.
We will therefore focus on couplings that are not present in the SM.

We are interested in the sensitivity to Higgs decays at the HL-LHC, where we expect about $N_h\sim 10^8$ Higgses to be produced
with $3$ ab$^{-1}$.  Estimating the SM Higgs branching ratios to the decays we consider, we find that they all have branching ratios larger than $10^{-8}$ so that all of these searches have a SM background.  Thus, looking at total decay rates, we should compare the new contribution to the fluctuations in the SM Higgs background.
\[
\eql{SMback}
\frac{\de\Ga_{\scr{O}}}{\Ga_h} N_h \gsim \left( 
\frac{\Ga_\text{SM}}{\Ga_h} N_h \right)^{1/2}.
\]
If this is satisfied, there is at least the possibility to distinguish the new contribution from the SM Higgs background.

We begin by considering the case where the interference between the BSM
and the SM contribution is negligible.
This may occur because the SM contribution is so small that the
BSM contribution dominates.
Another interesting case is where the BSM contribution is CP odd.
If the measurement performed is sufficiently inclusive that it
weights CP conjugate final states equally, the 
interference term between CP-even and CP-odd amplitudes cancels.
This occurs for example in the total rate summed over final state
spins.
Measurements of differential distributions may be sensitive to
interference terms, but these are beyond the simple estimates
performed here and should be studied on a case-by-case basis.  

We will estimate the size of the BSM contribution assuming that the
matrix element of the decay is constant, and that the decay is not
phase space suppressed.
The matrix elements for 2- and 3-body decays due to the insertion
of a BSM operator $\scr{O}$ are then approximated by
\begin{subequations}
\[
\scr{M}_{\scr{O}}(h \to 2) 
&\simeq \frac{c_{\scr{O}}}{v^{d_{\scr{O}} - 4}} 
m_h^{d_{\scr{O}} - 3},
\\
\scr{M}_{\scr{O}}(h \to 3) &\simeq \frac{c_{\scr{O}}}{v^{d_{\scr{O}} - 4}}
m_h^{d_{\scr{O}} - 4},
\]
\end{subequations}
where $d_{\scr{O}}$ is the dimension of the operator $\scr{O}$.
The corresponding decay rates are approximated by
\begin{subequations}
\[
\de\Ga_\scr{O}(h \to 2) &\simeq \frac{1}{16\pi m_h} 
\big| \scr{M}_{\scr{O}}(h \to 2) \big|^2
\simeq \frac{m_h}{16\pi} \big| c_\scr{O} \big|^2 
\left( \frac{m_h^2}{v^2} \right)^{d_{\scr{O}} - 4},
\\
\de\Ga_{\scr{O}}(h \to 3) &\simeq \frac{m_h}{512\pi^3}
\big| \scr{M}_{\scr{O}}(h \to 3) \big|^2
\simeq \frac{m_h}{512\pi^3} \big| c_\scr{O} \big|^2 
\left( \frac{m_h^2}{v^2} \right)^{d_{\scr{O}} - 4}
\]
\end{subequations}

To be of interest, we need to compare this deviation to the fluctuations in the SM Higgs background \Eq{SMback}, which is conservative since many of these will have additional backgrounds.  
This gives the bounds
\begin{subequations}
\[
\!\!\!
\text{2-body, no interference: }
|c_\scr{O}| &\gsim \big(4 \times 10^{-4}\big) 
\big(\text{BR}_{\text{SM}} \big)^{1/4}
\gap 2^{d_{\scr{O}} - 4}
\left( \frac{N_h}{10^8} \right)^{\!\! -1/4}\!\!\!,
\\
\!\!\!
\text{3-body, no interference: }
|c_{\scr{O}}| &\gsim \big(7 \times 10^{-3}\big) 
\big(\text{BR}_{\text{SM}} \big)^{1/4}
\gap 2^{d_{\scr{O}} - 4}
\left( \frac{N_h}{10^8} \right)^{\!\! -1/4}\!\!\!,
\]
\end{subequations}
where $\text{BR}_{\text{SM}}$ is the branching ratio of the
decay in the SM.  
The estimates for higher-dimension operators are more uncertain
due to the high powers of ratios of scales involved.

Now we consider the case where there is significant interference
with the SM.
In this case, we obtain a rough estimate by also approximating the SM
amplitude as a constant.
For example, for 2-body decays this gives
\[
\Ga_\text{SM}(h \to 2) \simeq \frac{1}{16\pi m_h} 
\big| \scr{M}_\text{SM}(h \to 2) \big|^2.
\]
The correction to the decay rate due to the BSM operator $\scr{O}$
is then
\[
\big| \de \Ga_{\scr{O}}(h \to 2) \big|
&\simeq \frac{1}{16\pi m_h} \big|\scr{M}_\text{SM} \big|
\big| \scr{M}_{\scr{O}} \big|
\nn
&\simeq
\frac{1}{16\pi} \Big[ 16\pi m_h \Ga_\text{SM}(h \to 2) \Big]^{1/2}
| c_\scr{O}| \left( \frac{m_h}{v} \right)^{d_\scr{O} - 4}.
\]
To be observable, the difference in the number of Higgs decays
compared with the SM must be larger than the fluctuations in
the SM background, as in \Eq{SMback}.
In this case, we find that the dependence on $\Ga_\text{SM}$ cancels
out in the bound, and we obtain the  bounds:
\begin{subequations}
\label{eq:interpheno}
\[
\text{2-body, interference: }
|c_\scr{O}| &\gsim \big(4 \times 10^{-6}\big)
\ggap 2^{d_{\scr{O}} - 4}
\left( \frac{N_h}{10^8} \right)^{-1/2},
\\
\text{3-body, interference: }
|c_{\scr{O}}| &\gsim \big(7 \times 10^{-5}\big)
\ggap 2^{d_{\scr{O}} - 4}
\left( \frac{N_h}{10^8} \right)^{-1/2}.
\]
\end{subequations}
Note that comparing to the no interference case, we see that when there is interference it allows better coupling sensitivity since we've estimated that $\text{BR}_{\text{SM}} \gtrsim 10^{-8}.$

These approximations made above are very crude, and are intended
only as a rough guide.
It will be interesting to compare them with detailed phenomenological
studies, but we leave that for future work.
In \sec{Results}, we will combine these estimates with the unitarity bounds
to identify some BSM operators that are worthy of further study.

\section{Results}
\label{sec:Results}
In this section, we present our results for the independent primary 
operators for the 3-point and 4-point amplitudes.
We do not consider flavor-violating operators and leave such generalizations for future work.
Equivalently, our results are presented for a single generation of
quarks and leptons.
This section consists mainly of the tables of operators,
with some brief comments in the main text.
We then use the results to discuss the most promising
primary observables for Higgs decays.

\subsection{3-Point Couplings}
We begin with the 3-point couplings.
These are equivalent to on-shell 3-point amplitudes
(for complex momenta), which have no Mandelstam invariants.
Therefore, all 3-point functions correspond to primary 
observables in our  terminology.
This problem has been previously studied by many authors, 
see for example \Refs{ Shadmi:2018xan,Durieux:2019eor,Helset:2020yio}.
Our main focus is the enumeration of the 4-particle observables,
but we have taken a fresh look at the 3-point functions to check
our approach.

The 3-point functions involving the Higgs boson are shown in Table~\ref{tab:h3p},
and the additional 3-point functions needed for Higgs processes that do not
involve the Higgs boson are shown in Table~\ref{tab:fv3p}.
The table gives the CP of the operator, a SMEFT
operator that contains the interaction, and the unitarity bound for
the coefficient of the operator, where the normalization for
the couplings is defined by \Eq{cO}.  %
\footnote{The SMEFT operator is included for comparison only;
we are not claiming that using the SMEFT operators in our
tables is a consistent EFT basis.}  
We have attempted to identify a SMEFT operator of lowest dimension, but this can be nontrivial and our method for determining it isn't systematic.  Finding lower dimension operators would weaken the bounds for unitarity constraints for large $E_\text{max}$.
To connect to the `Higgs basis' \cite{Falkowski:2001958}, we note that the interactions that appear at dimension 6 in SMEFT can be read off from our tables.
In some cases, we find that some four point functions in \cite{Falkowski:2001958} are redundant due to field redefinitions.

For the triple gauge boson couplings, we note that our approach
differs from the classic work \cite{Hagiwara:1986vm} in that
we are performing a systematic low-energy expansion of the kinematic
dependence.
As explained in \sec{onshell3} above, this necessarily involves an
interplay between 3-point and higher-point couplings.
We have put the effects of possible `form factors' of our 3-point
couplings into higher-point couplings.
Ref.~\cite{Hagiwara:1986vm} instead defines this in terms of form factors
whose momentum dependence must be specified to define a model
for experimental searches.
In particular, they include form factors for couplings of the
form $\scr{O}^{WWZ}_{4,5}$ with $Z_\mu$ replaced by 
$A_\mu$ 
even though these couplings are not $U(1)_\text{EM}$ gauge invariant.
(They restore gauge invariance by using a specific non-local form factor 
for these couplings that contains massless poles.)
We believe that our approach is more physically transparent and can
be systematically matched to EFT frameworks such as SMEFT.

\begin{table}[H]
\begin{center}
\footnotesize
\centering
\vspace{1 mm}
\renewcommand{\arraystretch}{1.2}
\tabcolsep6pt\begin{tabular}{|c|c|c|c|c|c|c|}
\hline
\multirow{2}{*}{$i$} & \multirow{2}{*}{$\scr{O}_i^{h \bar f f}$}   & \multirow{2}{*}{CP} & \multirow{2}{*}{$d_{\scr{O}_i}$}& SMEFT & $c$ Unitarity  \\[-5pt]
&   && & Operator & Bound \\
\hline
1 &  $ h \bar{\psi}_L  \psi_R +\hc$ & +&    \multirow{2}{*}{ 4}&   $H^\dagger H \bar{Q}_L \tilde H u_R +\hc$ &  \multirow{2}{*}{  $\frac{6}{E_\text{TeV}}, \frac{20}{E_\text{TeV}^2}$ }  \\
2 &  $  ih \bar{\psi}_L   \psi_R +\hc$&  $-$ & &  $iH^\dagger H \bar{Q}_L \tilde H u_R +\hc$& \\
\hline\hline
\multirow{2}{*}{$i$} & \multirow{2}{*}{$\scr{O}_i^{hZZ}$}   & \multirow{2}{*}{CP} &  \multirow{2}{*}{$d_{\scr{O}_i}$}& SMEFT & $c$ Unitarity  \\[-5pt]
&   & & & Operator & Bound \\
\hline
1 &  $  h Z_\mu Z^\mu$&  +  &  \multirow{1}{*}{ 3}&  $H^\dagger H D^\mu H^\dagger D_\mu H$ &  \multirow{1}{*}{ $\frac{0.2}{E_\text{TeV}^2}$ }\\
2 &  $  h Z_{\mu\nu} Z^{\mu\nu}$& + &  \multirow{2}{*}{ 5}&$H^\dagger H W^a_{\mu\nu} W^{a \mu\nu}$   &  \multirow{2}{*}{  $\frac{2}{E_\text{TeV}^2}$}  \\
3 & $  h Z_{\mu\nu} \widetilde{Z}^{\mu\nu}$&  $-$ &  &$H^\dagger H W^a_{\mu\nu} \widetilde{W}^{a \mu\nu}$   & \\
\hline\hline
\multirow{2}{*}{$i$} & \multirow{2}{*}{$\scr{O}_i^{h WW}$}   & \multirow{2}{*}{CP} &  \multirow{2}{*}{$d_{\scr{O}_i}$}& SMEFT & $c$ Unitarity  \\[-5pt]
&   & & & Operator & Bound \\
\hline
1&  $  h W^+_\mu W^{-\mu}$& +  & \multirow{1}{*}{ 3} &  $H^\dagger H D^\mu H^\dagger D_\mu H$ &   \multirow{1}{*}{ $\frac{0.2}{E_\text{TeV}^2}$ } \\
2 &  $  h W^+_{\mu\nu} W^{-\mu\nu}$&  + & \multirow{2}{*}{ 5} &$H^\dagger H W^a_{\mu\nu} W^{a \mu\nu}$   &  \multirow{2}{*}{  $\frac{2}{E_\text{TeV}^2}$} \\
3 & $  h W^+_{\mu\nu} \widetilde{W}^{\mu\nu}$& $-$ & &$H^\dagger H W^a_{\mu\nu} \widetilde{W}^{a \mu\nu}$   & \\
\hline\hline
\multirow{2}{*}{$i$} & \multirow{2}{*}{$\scr{O}_i^{h Z \ga}$}   & \multirow{2}{*}{CP} &  \multirow{2}{*}{$d_{\scr{O}_i}$}& SMEFT & $c$ Unitarity  \\[-5pt]
&   & & & Operator & Bound \\
\hline
1 &  $  h F_{\mu\nu} Z^{\mu\nu}$& + & \multirow{2}{*}{ 5}&$H^\dagger \si^a H W^a_{\mu\nu} B^{ \mu\nu}$ &   \multirow{2}{*}{  $\frac{2}{E_\text{TeV}^2}$}  \\
2& $  h F_{\mu\nu} \widetilde{Z}^{\mu\nu}$&  $-$ &  &$H^\dagger \si^a H B_{\mu\nu} \widetilde{W}^{a \mu\nu}$   &   \\
\hline\hline
\multirow{2}{*}{$i$} & \multirow{2}{*}{$\scr{O}_i^{h \ga \ga}$}   & \multirow{2}{*}{CP} &  \multirow{2}{*}{$d_{\scr{O}_i}$}& SMEFT & $c$ Unitarity  \\[-5pt]
&   & & & Operator & Bound \\
\hline
1&  $  h F_{\mu\nu} F^{\mu\nu}$& + & \multirow{2}{*}{ 5}&$H^\dagger H B_{\mu\nu} B^{ \mu\nu}$   &   \multirow{2}{*}{  $\frac{2}{E_\text{TeV}^2}$}    \\
2 & $  h F_{\mu\nu} \widetilde{F}^{\mu\nu}$&  $-$  &&$H^\dagger H B_{\mu\nu} \widetilde{B}^{ \mu\nu}$   & \\
\hline\hline
\multirow{2}{*}{$i$} & \multirow{2}{*}{$\scr{O}_i^{h GG}$}   & \multirow{2}{*}{CP} & \multirow{2}{*}{$d_{\scr{O}_i}$}& SMEFT & $c$ Unitarity  \\[-5pt]
&   && &Operator & Bound\\
\hline
1&  $  h G_{\mu\nu} G^{\mu\nu}$& + &  \multirow{2}{*}{ 5}&$H^\dagger H G_{\mu\nu} G^{\mu\nu}$   &   \multirow{2}{*}{ $\frac{2}{E_\text{TeV}^2}$ } \\
2 & $  h G_{\mu\nu} \widetilde{G}^{\mu\nu}$& $-$ & &$H^\dagger H G_{\mu\nu} \widetilde{G}^{\mu\nu}$   & \\
\hline\hline
\multirow{2}{*}{} & \multirow{2}{*}{$\scr{O}^{h hh}$}   & \multirow{2}{*}{CP} &  \multirow{2}{*}{$d_{\scr{O}_i}$} & SMEFT & $c$ Unitarity  \\[-5pt]
&   &&& Operator & Bound \\
\hline
 & $  hhh $ & +  & 3& $|H|^6$   & $\frac{80}{E_\text{TeV}}, \frac{200}{E_\text{TeV}^2}$ 
 \\[4pt]
\hline
\end{tabular}
\begin{minipage}{5.7in}
\medskip
\caption{\small 
Three point functions that involve the Higgs boson.  
We write $V_{\mu\nu} = \partial_\mu V_\nu - \partial_\nu V_\mu$ 
and $\widetilde{V}_{\mu\nu} = \frac{1}{2}\epsilon_{\mu\nu\rho\si}V^{\rho\si}$
for $V = W, Z$, while $F_{\mu\nu}$ and $G_{\mu\nu}$ are the field strength
tensors for the photon and gluon, respectively. 
We have omitted the color indices of the gluon fields.
The last column gives the maximum allowed value for the coupling $c$
defined in \Eq{cO} allowed by tree-level unitarity,
where $E_\text{TeV}$ is the unitarity violating 
scale in units of TeV.
\label{tab:h3p}}
\end{minipage}
\end{center}
\end{table}

\newpage
\begin{table}[H]
\begin{center}
\footnotesize
\centering
\vspace{1 mm}
\renewcommand{\arraystretch}{1.2}
\tabcolsep6pt\begin{tabular}{|c|c|c|c|c|c|}
\hline
\multirow{2}{*}{$i$} & \multirow{2}{*}{$\scr{O}_i^{\ga \bar {f}  f}$}  & \multirow{2}{*}{CP} & \multirow{2}{*}{$d_{\scr{O}_i}$} & SMEFT & $c$ Unitarity  \\[-5pt]
&   && & Operator & Bound\\
\hline
1&  $  F^{\mu\nu} \bar{\psi}_L \si_{\mu\nu}   \psi_R +\hc  $& + &  \multirow{2}{*}{5}&$B_{\mu\nu} \bar{Q}_L\si^{\mu\nu} \tilde H u_R +\hc$   &  \multirow{2}{*}{$\frac{2}{E_\text{TeV}^2}$}  \\
2&  $   i F^{\mu\nu}  \bar{\psi}_L \si_{\mu\nu}   \psi_R +\hc$  &  $-$& &$iB_{\mu\nu} \bar{Q}_L\si^{\mu\nu} \tilde H u_R +\hc$    & \\
\hline\hline
\multirow{2}{*}{$i$} & $\scr{O}_i^{Z\bar f f}$  & \multirow{2}{*}{CP} &  \multirow{2}{*}{$d_{\scr{O}_i}$} &SMEFT & $c$ Unitarity  \\[-5pt]
&  ($Z\to W$ gives $W\bar{f}f'$) && &Operator & Bound\\
\hline
1 & $  Z^{\mu} \bar{\psi}_L\ga_{\mu} \psi_L $&  +  &  \multirow{2}{*}{4}&$ i  H^\dagger  \overset\leftrightarrow{D}_\mu  H  \bar Q_L \ga^\mu Q_L$   &  \multirow{2}{*}{$\frac{0.6}{E_\text{TeV}^2}$} \\
2 &  $  Z^{\mu} \bar{\psi}_R \ga_{\mu} \psi_R $& + & &$ i  H^\dagger   \overset\leftrightarrow{D}_\mu  H \bar u_R \ga^\mu u_R$   &    \\
\hline
3 & $  Z^{\mu\nu} \bar{\psi}_L  \si_{\mu\nu} \psi_R +\hc $& + & \multirow{2}{*}{5}  &$W^a_{\mu\nu} \bar{Q}_L\si^{\mu\nu} \si^a \tilde H u_R +\hc$   & \multirow{2}{*}{$\frac{2}{E_\text{TeV}^2}$} \\
4 &  $  Z^{\mu\nu} i \bar{\psi}_L \si_{\mu\nu} \psi_R +\hc$& $-$& &$iW^a_{\mu\nu} \bar{Q}_L\si^{\mu\nu} \si^a \tilde H u_R +\hc$   &    \\
\hline\hline
\multirow{2}{*}{$i$} & \multirow{2}{*}{$\scr{O}_i^{WW Z}$}  & \multirow{2}{*}{CP} &  \multirow{2}{*}{$d_{\scr{O}_i}$} &SMEFT & $c$ Unitarity  \\[-5pt]
&   && &Operator & Bound\\
\hline
1& $iW^+_{\mu\nu}W^{-\, \mu}Z^\nu +\hc$ & $+$& \multirow{3}{*}{4}& $i H^\dagger \si^a \overset\leftrightarrow{D^\mu}  H D^\nu W^a_{\mu\nu} $& \multirow{3}{*}{$\frac{0.5}{E_\text{TeV}^2}$}\\
2&$i  W^+_{\mu} W^-_{\nu}Z^{\mu\nu} $ &  $+$ && $i (D_\mu H)^\dagger \si^a D_\nu H W^{a\mu\nu}$ & \\
3&$ i  W^+_\mu W^-_\nu \widetilde{Z}^{\mu\nu}$ &  $-$ && $i (D_\mu H)^\dagger \si^a D_\nu H \widetilde{W}^{a\mu\nu}$   & \\
\hline
4& $-  W^+_\mu W^-_\nu (\partial^\mu Z^\nu + \partial^\nu Z^\mu)$ &  $-$ & 4& $i D_\mu H^\dagger D_\nu H H^\dagger D^{\mu \nu} H +\hc$ & $\frac{4 \cdot 10^{-3}}{E_\text{TeV}^4}$ \\[4pt]
\hline
5& $ \epsilon^{\mu\nu\rho\si}(W^+_\mu \overset\leftrightarrow{D}_\rho W^-_{\nu})Z_\si$ & $+$ & 4&  $ \epsilon^{\mu\nu\rho\si}(H^\dagger \si^a D_\mu H ) W^{a}_{\rho\nu} H^\dagger D_\si H +\hc$ & $\frac{0.04}{E_\text{TeV}^3}$, 
$\frac{0.1}{E_\text{TeV}^4}$ \\[4pt]
\hline
6& $  i W^+_{\lambda\mu}W^{-\, \mu}_{\quad \nu} Z^{\nu\lambda} $ & $+$ & \multirow{2}{*}{6}&  $   \epsilon_{abc}W^a_{\lambda\mu}W^{b\, \mu}_{\quad \nu} W^{c\nu\lambda} $ & \multirow{2}{*}{$\frac{5}{E_\text{TeV}^2}$} \\
7& $  i W^+_{\lambda\mu}W^{-\, \mu}_{\quad \nu} \widetilde{Z}^{\nu\lambda} $ & $-$ & &  $  \epsilon_{abc}W^a_{\lambda\mu}W^{b\, \mu}_{\quad \nu} \widetilde{W}^{c\nu\lambda} $ & \\
\hline\hline
\multirow{2}{*}{$i$} & \multirow{2}{*}{$\scr{O}_i^{WW\ga}$}  & \multirow{2}{*}{CP} &  \multirow{2}{*}{$d_{\scr{O}_i}$} &SMEFT & $c$ Unitarity  \\[-5pt]
&   && & Operator & Bound\\
\hline
1& $i W^+_{\mu}W^{-}_{\nu}F^{\mu\nu} $ & $+$ &  \multirow{2}{*}{4}& $i (D_\mu H)^\dagger D_\nu H B^{\mu\nu}$ & \multirow{2}{*}{$\frac{0.5}{E_\text{TeV}^2}$} \\
2&$i W^+_{\mu}W^{-}_{\nu}\widetilde{F}^{\mu\nu}$ &  $-$ && $i (D_\mu H)^\dagger D_\nu H \widetilde{B}^{\mu\nu}$     & \\
\hline
3& $i W^+_{\mu\nu}W^{-\, \nu \rho}F^{\; \mu}_{\rho} $ & $+  $ &\multirow{2}{*}{6} & $\epsilon_{abc} W^a_{\mu\nu}W^{b\, \nu \rho}W^{c\, \mu}_{\rho} $ & \multirow{2}{*}{$\frac{5}{E_\text{TeV}^2}$}\\
4&$i W^+_{\mu\nu}W^{-\, \nu \rho}\widetilde{F}^{\; \mu}_{\rho}$ &  $-$ && $\epsilon_{abc} W^a_{\mu\nu}W^{b\, \nu \rho}\widetilde{W}^{c\, \mu}_{\rho} $  & \\
\hline\hline
\multirow{2}{*}{$i$} & \multirow{2}{*}{$\scr{O}_i^{ggg}$}  & \multirow{2}{*}{CP} & \multirow{2}{*}{$d_{\scr{O}_i}$} & SMEFT & $c$ Unitarity  \\[-5pt]
&   &&& Operator & Bound\\
\hline
1& $f_{ABC} G^A_{\mu\nu}G^{B\, \nu \rho}G^{C\, \mu}_{\rho} $ & $+$ & \multirow{2}{*}{6}& $f_{ABC} G^A_{\mu\nu}G^{B\gap \nu \rho}G^{C\gap \mu}_{\rho} $ & \multirow{2}{*}{$\frac{2}{E_\text{TeV}^2}$}\\
2&$f_{ABC} G^A_{\mu\nu}G^{B\gap \nu \rho}\widetilde{G}^{C\gap \mu}_{\rho} $ &  $-$ && $f_{ABC} G^A_{\mu\nu}G^{B\gap \nu \rho}\widetilde{G}^{C\gap \mu}_{\rho} $  & \\
\hline
\end{tabular}
\begin{minipage}{5.7in}
\medskip
\caption{\small
Additional three 
point functions needed to calculate 4-point amplitudes involving the Higgs. 
The notation is the same as in Table~\ref{tab:h3p}.
Here  $\si^{\mu\nu} =  \frac i 4 [\ga^\mu, \ga^\nu]$, $\si^a$ are Pauli matrices, 
and $H^\dagger \overset\leftrightarrow{D}_\mu  H = H^\dagger D_\mu H - (D_\mu H)^\dagger \, H$.  
\label{tab:fv3p}}
\end{minipage}
\end{center}
\end{table}

\newpage
\subsection{4-Point Couplings}
Our results for 4-point operators are summarized in 
Tables~\ref{tab:hZff}--\ref{tab:hhhV}.
The notation is hopefully self-explanatory;
to save space, we have used $\d_{\mu\nu\rho} = \d_\mu \d_\nu \d_\rho$, $D_{\mu\nu\rho} = D_\mu \d_\nu \d_\rho$,
{\it etc\/}.
There are several cases for which we do not provide separate tables,
because the operators can be read off from other tables by simple
substitutions:
\begin{itemize}
\item
$hW\bar{f}f'$ can be
read off from $hZ\bar{f}f$ in Table~\ref{tab:hZff}
with the substitution $Z_\mu\bar{f}f \to W_\mu \bar{f} f'$
and $\d_\mu \to D_\mu$.
\item
$hg\bar{f}f$ can be obtained from $h\ga\bar{f}f$
in the bottom part of Table~\ref{tab:hZff} 
with the substitution $F_{\mu\nu} \to G_{\mu\nu}^A T_A$.
The operators are $SU(3)_\text{C}$ gauge invariant only if the fermions 
are quarks.
\item
$hZgg$ can be obtained from $hZ\ga\ga$ in 
Table~\ref{tab:hZgamgamsym} with the substitution
$F_{\mu\nu} F_{\rho\si} \to G_{\mu\nu}^A G_{\rho\si}^A$.
\item
$hhZZ$ can be obtained from $hhWW$ in 
Table~\ref{tab:hhVV}
by replacing $W_\mu \to Z_\mu$.
When this is done, the operators numbered 5, 7 and 8 vanish
by symmetry, so there are only 6 nonzero operators in this case.
\item
For $hhhh$ the only primary operator  is $h^4$,
and we have not made a table for that.
\end{itemize}

There are other cases where the results are closely related,
but additional corrections must be made.
For example, we can take operators involving $Z$ and convert them to operators with a photon, by taking $Z_
\mu \to A_\mu$ and forming the field strength for the photon by using derivatives and anti-symmetrizing.  This allows $hZ\bar{f}f, hZgg$ to be respectively converted to $h\gamma \bar{f}f$ and $h\ga gg$.

The tables list the primary operators.
In the on-shell amplitude language, the remaining amplitudes are obtained by
multiplying each operator by a power series in the Mandelstam variables.
In the operator language, these correspond to operators with additional
derivatives with the Lorentz indices contracted between them.
For operators where all particles are distinguishable, this is
simply a series in the Mandelstam variables $s$, $t$, and $u$
(with $s + t + u$ fixed).
For operators with identical particles, these additional terms must
be appropriately symmetrized.
For $hZ\ga\ga$, $h\ga gg$ (Table~\ref{tab:hZgamgamsym}),
$hZZ\ga$ (Table~\ref{tab:hZZgamma}),
$hh\bar{f}f$ (Table~\ref{tab:hhff}),
$hhWW$, $hhZ\ga$ and $hh\ga\ga$ (Table~\ref{tab:hhVV}),
we can add arbitrary powers of
$s$ and $(t - u)^2$.
For $h\ga\ga\ga$ (Table~\ref{tab:hgamgamgam}),
$hggg$ (Table~\ref{tab:hggg}),
$hZZZ$ (Table~\ref{tab:hZZZ}),
$hhhZ$ and $hhh\ga$ (Table~\ref{tab:hhhV}),
we can add arbitrary powers of
$s^2 + t^2 + u^2$ and $stu$.
As an example, adding a factor of $s^2+t^2+u^2$ to $h \d^{\mu} Z^{\nu} Z_{\mu} Z_{\nu}$ can be done by adding four derivatives, {\it i.e.}~$h \d^{\mu} Z^{\nu} \partial^\alpha \partial^\beta Z_{\mu} \partial_\alpha \partial_\beta Z_{\nu}$.

The tables give unitarity bounds on the coefficients of the operators
(see \sec{Unitarity}).
As one might expect, the unitarity bounds become more stringent for
operators of higher dimension.
These bounds should be used only as a very rough guide, especially 
for the operators with high mass dimension.

Our final results are in full agreement
with the Hilbert series counting in all cases (see \sec{HilbertSeries}).
We also agree with the results of Ref.~\cite{Durieux:2020gip} 
in all cases where they overlap.
We found a discrepancy in the results for $hWWZ$ 
(see Table~\ref{tab:hWWZ}) in an earlier version
of their paper, but our results agree after they identified and corrected
a mistake.
Our results also include massless particles, the effects of symmetrization 
for identical particles, and we have found all primary operators
to arbitrarily high dimension, at least in the cases where there are
two or fewer identical particles (see \sec{sec:enumerating}).

\subsection{Primary Observables for Higgs Decay}
\scl{PrimaryDecays}
We now use the results in the tables to identify promising primary observables
to search for new physics in Higgs decays.
We limit ourselves to CP even operators,
so that it is clear that there is interference with SM processes.
(Also, CP-odd new physics effects may be suppressed by approximate
CP symmetry.)
In this case \Eq{interpheno} gives an estimate for the minimal value of the
coefficients in order for the new contribution to the decay to be observable
at the HL-LHC.
We compare this to the bound on the coefficient arising from the unitarity
bounds in the tables.

In this way, we find that the following operators are potentially observable
at the LHC assuming a unitarity violating scale above 10~TeV:
\[
\scr{O}_{1}^{h \bar f f},
\quad \scr{O}^{hVV}_{1,2},
\quad \scr{O}^{hZ\ga}_{1},
\quad \scr{O}^{h\ga\ga}_{1},
\quad \scr{O}^{hGG}_{1},
\quad \scr{O}^{hZ \bar f f}_{1,2,3},
\quad\scr{O}^{hW \bar f f'}_{1,2,3}, 
\quad \scr{O}^{h\ga \bar f  f}_{1},
\]
where $V = W,Z$.
The next class of operators are those that are potentially observable with
a unitarity violating scale between 1 and 10~TeV:
\[
\scr{O}^{hZ \bar f f}_{5,7,9,11},
\quad \scr{O}^{hW \bar f f'}_{5,7,9,11}, 
\quad \scr{O}^{h\ga \bar f  f}_{3,5,7}, 
\quad \scr{O}_{1,4,5,8,9}^{hZ\ga\ga}, 
\quad \scr{O}_{1,2}^{h\ga g g}, 
\quad \scr{O}_{1,3}^{hg g g,f}.
\]
These are also interesting, but it may be that new physics models
that can give these effects can be better probed by direct searches for new
heavy particles.
The remaining operators are observable only if the unitarity violating
scale is below 1 TeV:
\[
\scr{O}^{hZ\ga\ga}_{11}, 
\quad \scr{O}_{5,6}^{h\ga g g}, 
\quad \scr{O}_{1,3,4,7}^{h\ga\ga\ga}, 
\quad \scr{O}_{1,3,4,7}^{h g g g,d}, 
\quad \scr{O}_{5,7}^{hg g g,f}.
\]
These are presumably already constrained, and not as theoretically motivated
as the others.

We see that there are a large number of observables that worthy of further
investigation.
This motivates searches for BSM effects in Higgs 2-body decays,
as well as 3-body decays to 
$Z\bar{f}f$, 
$W\bar{f}f'$, 
$\ga \bar{f}f$, 
$g \bar{f}f$, 
$Z\ga\ga$, 
$\ga gg$, 
and $ggg$.   
The decays to strongly-interacting particles are likely very challenging
due to QCD backgrounds that we have neglected.
We note that some detailed phenomenological studies on the effects of higher-dimension
operators on 3-body decays have already been performed.
For example, \cite{Boselli:2017pef} considers effects equivalent to 
some of the operators above in the decay 
$h \rightarrow e^- e^+ \mu^- \mu^+$, but not all of them.
We leave further detailed study of these effects for future work.

\newpage

\begin{table}[H]
\small
\centering
\vspace{1 mm}
\renewcommand{\arraystretch}{1.2}
\tabcolsep6pt\begin{tabular}{|c|c|c|c|c|c|c|}
\hline
\multirow{2}{*}{$i$} & \multirow{2}{*}{${\cal{O}}_i^{hZ\bar{f}f}$}    & \multirow{2}{*}{CP} & \multirow{2}{*}{$d_{\scr{O}_i}$} &SMEFT &  $c$ Unitarity  \\[-5pt]
& &  & &Operator  & Bound\\
\hline
1 &  $ h Z^\mu \bar{\psi}_L\ga_\mu  \psi_L$& +&  \multirow{2}{*}{5} &   $ i  \big(H^\dagger  \overset\leftrightarrow{D}_\mu  H\big)  \bar Q_L \ga^\mu Q_L$ &  \multirow{2}{*}{$\frac{0.6}{E_\text{TeV}^2}, \frac{5}{E_\text{TeV}^4}$}  \\
2 &  $  h Z^\mu \bar{\psi}_R\ga_\mu \psi_R$&  +  & &  $ i  \big( H^\dagger   \overset\leftrightarrow{D}_\mu  H \big) \bar u_R \ga^\mu u_R$ & \\
\hline
3 &  $  h Z^{\mu\nu} \bar{\psi}_L\si_{\mu\nu}  \psi_R + \hc$& + &  \multirow{2}{*}{6}&  $\bar{Q}_L \si^{\mu\nu}u_R  \si^a\widetilde{H} W^a_{\mu\nu} + \hc$ &   \multirow{2}{*}{ $\frac{2}{E_\text{TeV}^2}, \frac{10}{E_\text{TeV}^4}$}   \\
4 &  $ i h \widetilde{Z}_{\mu\nu}  \bar{\psi}_L \si^{\mu\nu}   \psi_R + \hc$&  $-$ & &   $i\bar{Q}_L \si^{\mu\nu}u_R  \si^a\widetilde{H} \widetilde{W}^a_{\mu\nu} + \hc$ &\\
\hline
5 &  $i h Z^\mu \big(\bar{\psi}_L \overset\leftrightarrow{\d}{}_\mu \psi_R \big) + \hc$& $+$ &  \multirow{4}{*}{6}&  $  \big( H^\dagger  \overset\leftrightarrow{D}_\mu H\big)  \big(\bar{Q}_L  \overset\leftrightarrow{D}{}^\mu u_R\big) \widetilde{H} + \hc$ &  \multirow{4}{*}{ $\frac{0.1}{E_\text{TeV}^3}, \frac{4}{E_\text{TeV}^6}$} \\
6 &$ h Z^\mu \d_\mu\!\big(\bar{\psi}_L  \psi_R\big) + \hc$ & $-$ &  &  
$i\big( H^\dagger  \overset\leftrightarrow{D}_\mu  H\big)  D^\mu \big(\bar{Q}_L   u_R\big) \widetilde{H}+\hc$ &\\
\hline
\end{tabular}
\begin{minipage}{5.7in}
\medskip
\caption{\small Primary operators for couplings of the form $hZ\bar{f}f$ and $h\ga f \bar{f}$.  
As noted in the text, the $hW\bar{f}f'$ operators can be obtained
from the $hZ\bar{f}f$ operators
by the replacement $Z\bar{f}f \to W\bar{f}f'$,
and the $hg\bar{f}f$ operators can be obtained from the
$h\ga\bar{f}f$ operators by the replacement
$F_{\mu\nu} \to G_{\mu\nu}$.
\label{tab:hZff}}
\end{minipage}
\end{table}

\begin{table}[H]
\scriptsize
\begin{center}
\renewcommand{\arraystretch}{1.4}
\tabcolsep5pt\begin{tabular}{|c|c|c|c|c|c|c|}
\hline
 \multirow{2}{*}{$i$} &   \multirow{2}{*}{${\cal{O}}_i^{hZ\ga\ga}$}   & \multirow{2}{*}{CP} & \multirow{2}{*}{$d_{\scr{O}_i}$} & SMEFT & $c$ Unitarity \\[-5pt]
     & 
     &  & &Operator & Bound\\
\hline
 1 & $ h\d^{\mu} F^{\al\be} \widetilde{F}_{\alpha\beta}Z_{\mu}$ &  $+$ &\multirow{2}{*}{7}&  $ i  H^\dagger  \overset\leftrightarrow{D}_\mu H\d^{\mu} B^{\al\be} \widetilde{B}_{\alpha\beta}$ &    \multirow{2}{*}{$\frac{0.1}{E_\text{TeV}^3}$}\\
 2 & $h \d^{\mu}F^{\nu\rho}F_{\mu\nu}Z_\rho $ &  $-$ &&  $ i  H^\dagger  \overset\leftrightarrow{D}_\rho H  \d^{\mu}B^{\nu\rho}B_{\mu\nu} $&  \\
\hline
 3& $h F^{\mu\nu}F_{\mu\rho}\d_\nu Z^\rho $ &  $-$ &7&  $i H^\dagger D_\nu D^\rho H B^{\mu\nu}B_{\mu\rho} +\hc$ & $\frac{0.1}{E_\text{TeV}^3}, \frac{1}{E_\text{TeV}^5}$
 \\[4pt]
\hline
 4& $ h \d_\mu F^{\al\nu} F^{\mu\de} \d_\nu  \widetilde{Z}^{\alpha \delta}$ & $+$ & \multirow{2}{*}{9}&  $\ H^\dagger 
 \si^a H \d_\mu B^{\al\nu} B^{\mu\de} D_\nu  \widetilde{W}^{a\alpha \delta}$ &  \multirow{2}{*}{$\frac{0.02}{E_\text{TeV}^5}, \frac{0.07}{E_\text{TeV}^6}$} \\
 5 & $h F^{\al\mu} \overset\leftrightarrow{\d}{}^{\nu}\widetilde{ F}_{\alpha\beta} \d_\nu  Z_\mu^{\ \be}$ &  $+$& &  $ H^\dagger 
 \si^a H B^{\al\mu} \overset\leftrightarrow{\d}{}^{\nu} \widetilde{B}_{\alpha \beta} D_\nu  W_\mu^{a\be}$ & \\
\hline
 6 & $ h F^{\mu\nu}\overset\leftrightarrow{\d}{}^{\si} 
 F_{\mu\rho}\d_\si \d_\nu  Z^\rho $ &  $-$ & \multirow{2}{*}{9}&  $ i H^\dagger D_{\si \nu \rho} H B^{\mu\nu}\overset\leftrightarrow{\d}{}^{\si}
 B_{\mu\rho} + \hc $  &  \multirow{2}{*}{$\frac{2\cdot 10^{-3}}{E_\text{TeV}^6}, \frac{0.02}{E_\text{TeV}^8}$}\\
\hline
\end{tabular}
\begin{minipage}{5.7in}
\medskip
\caption{\small Primary operators for  $hZ\ga\ga$. 
As noted in the text, the $hZgg$ operators can be
obtained from these by the replacement $F_{\mu\nu} \to G_{\mu\nu}$.
\label{tab:hZgamgamsym}}
\end{minipage}
\end{center}
\end{table}

\begin{table}[H]
\scriptsize
\begin{adjustwidth}{-.5in}{-.5in}  
\begin{center}
\renewcommand{\arraystretch}{1.4}
\tabcolsep8pt\begin{tabular}{|c|c|c|c|c|c|c|}
\hline
 \multirow{2}{*}{$i$} &  \multirow{2}{*}{${\cal{O}}_i^{h \ga g g}$}   & \multirow{2}{*}{CP}& \multirow{2}{*}{$d_{\scr{O}_i}$}  & SMEFT & $c$ Unitarity \\[-5pt]
     &    & & & Operator & Bound\\
\hline
1 & $ hD_{\mu} G^{\al\nu} G^{\mu\de}\d_\nu \widetilde{F}_{\alpha \delta}$  &  $+$ &  \multirow{4}{*}{9}&  $H^\dagger H \d_{\mu} G^{\al\nu} G^{ \mu\de}\d_\nu \widetilde{B}_{\alpha \delta}$  &   \multirow{4}{*}{$\frac{0.02}{E_\text{TeV}^5}, \frac{0.07}{E_\text{TeV}^{6}}$}  \\
2 & $ h G^{\al\mu} \overset\leftrightarrow{D}_\nu \widetilde{G}_{ \alpha \beta}\d^\nu F_{\mu}^{\; \be}$  & $+$ & & $  H^\dagger H G^{ \al\mu} \overset\leftrightarrow{D}_\nu \widetilde{G}_{\alpha \beta}\d^\nu B_{\mu}^{\; \be}$ &  \\
3 & $ h D_{\si} G^{\mu\nu} G_{ \mu\rho}\d_\nu F^{\si\rho}$  & $-$ && $ H^\dagger H D_{\si} G^{\mu\nu} G_{ \mu\rho}\d_\nu B^{\si\rho}$ &  \\
4 & $ h  G^{\mu\nu} D_\si G_{ \mu\rho}\d_\nu F^{\si\rho}$  &  $-$& & $ H^\dagger H  G^{ \mu\nu} D_\si G_{ \mu\rho}\d_\nu B^{\si\rho}$ &  \\
\hline
5 & $ hD_{\si} G^{ \al\mu}  \overset\leftrightarrow{D}_\nu \widetilde{G}_{\alpha\beta}\d^\nu_{\ \mu} F^{\si \be}$  &  $+$ & \multirow{4}{*}{11} & $H^\dagger H D_{\si} G^{\al\mu}  \overset\leftrightarrow{D}_\nu \widetilde{G}_{\alpha \beta}\d^\nu_{\ \mu} B^{\si \be}$  &    \multirow{4}{*}{$\frac{1 \cdot 10^{-3}}{E_\text{TeV}^7}, \frac{4 \cdot 10^{-3}}{E_\text{TeV}^{8}}$}\\
6 & $ hD_{\mu} G^{\al\be}  \overset\leftrightarrow{D}_\nu D_{\si } \widetilde{G}_{\alpha\beta}\d^\nu  F^{\si \mu}$  &  $+$& & $ H^\dagger H D_{\mu} G^{\al\be}  \overset\leftrightarrow{D}_\nu D_{\si } \widetilde{G}_{\alpha\beta}\d^\nu  B^{\si \mu}$  &  \\
\hline
\end{tabular} 
\end{center}
\end{adjustwidth}
\caption{\small Primary operators for $h\ga gg$ and $h\ga\ga\ga$.}
\label{tab:hgamgamgam}
\end{table}

\begin{table}[H]
\scriptsize
\begin{adjustwidth}{-.5in}{-.5in} 
\begin{center}
\renewcommand{\arraystretch}{1.4}
\tabcolsep10pt\begin{tabular}{|c|c|c|c|c|c|}
\hline
 \multirow{2}{*}{$i$} & ${\cal{O}}_i^{h ggg, d}$   & \multirow{2}{*}{CP} & \multirow{2}{*}{$d_{\scr{O}_i}$} & SMEFT & $c$ Unitarity \\[-5pt]
    &  $\!\!\!\!$Color indices contracted with $d_{ABC}\!\!\!\!\!$  & & & Operator & Bound\\
\hline
$1$ & $ hD_{\si} G^{\al\mu}  \overset\leftrightarrow{D}_\nu \widetilde{G}_{\alpha\beta} D^\nu_{\ \mu} G^{\si \be}$   & $+$ &  \multirow{2}{*}{11}&$ H^\dagger H D_{\si} G^{ \al\mu}  \overset\leftrightarrow{D}_\nu \widetilde{G}_{\alpha\beta}  D^\nu_{\ \mu} G^{\si \be}$  &  \multirow{2}{*}{$\frac{1 \cdot 10^{-3}}{E_\text{TeV}^7}, \frac{4 \cdot 10^{-3}}{E_\text{TeV}^8}$}  \\
$2$ & $ hD_{\ga} G^{\mu\nu} \overset\leftrightarrow{D}_\si G_{ \mu\rho} D^\si_{\ \nu} G^{\ga\rho}$  &  $-$& &$ H^\dagger H D_{\ga} G^{ \mu\nu} \overset\leftrightarrow{D}_\si G_{ \mu\rho} D^\si_{\ \nu} G^{\ga\rho}$    &  \\
\hline
$3$ & $hD_{\mu\rho\si} G^{\al\nu} D^{\rho\si} G^{\mu\de}D_\nu \widetilde{G}_{\alpha\delta}$  &  $+$ & \multirow{4}{*}{13}& $ H^\dagger H D_{\mu\rho\si} G^{\al\nu} D^{\rho\si} G^{\mu\de} D_\nu \widetilde{G}_{\alpha\delta}$   & \multirow{4}{*}{$\frac{8\cdot 10^{-5}}{E_\text{TeV}^9}, \frac{3\cdot 10^{-4}}{E_\text{TeV}^{10}}$} \\
$4$ & $ h D_{\rho\si} G^{\al\mu} \overset\leftrightarrow{D}_\nu D^{\rho\si} \widetilde{G}_{\alpha\beta}D^\nu G_{\mu}^{\ \be}$  &  $+$ && $H^\dagger H D_{\rho\si} G^{\al\mu} \overset\leftrightarrow{D}_\nu D^{\rho\si} \widetilde{G}_{\alpha\beta} D^\nu G_{\mu}^{\ \be}$ &  \\
$5$ & $ hD_{\si}^{\al\be}  G^{\mu\nu} D_{\al\be} G_{\mu\rho}D_\nu G^{\si\rho}$  &  $-$ & &  $ H^\dagger H D_{\si}^{\ \al\be} G^{\mu\nu} D_{\al\be} G_{\mu\rho}D_\nu G^{\si\rho}$ &  \\
$6$ & $ h  D^{\al\be} G^{\mu\nu}  D_{\si\al\be} G_{\mu\rho}D_\nu G^{\si\rho}$  &  $-$ && $ H^\dagger H  D^{\al\be} G^{\mu\nu}  D_{\si\al\be} G_{\mu\rho} D_\nu G^{\si\rho}$ &  \\
\hline 
\end{tabular}
\end{center}
\end{adjustwidth}
\caption{\small Primary operators for $hggg$.
Gluon color indices are omitted.}
\label{tab:hggg}
\end{table}

\begin{table}[H]
\scriptsize
\begin{adjustwidth}{-.5in}{-.5in} 
\begin{center}
\renewcommand{\arraystretch}{1.4}
\tabcolsep6pt\begin{tabular}{|c|c|c|c|c|c|}
\hline
 \multirow{2}{*}{$i$} &  \multirow{2}{*}{${\cal{O}}_i^{h WW\ga}$}   & \multirow{2}{*}{CP}& \multirow{2}{*}{$d_{\scr{O}_i}$}  & SMEFT & $c$ Unitarity \\[-5pt]
     &    & & &Operator & Bound\\
\hline
1& $i h W^+_\mu W^-_\nu F^{\mu\nu}+ \hc$ & $+$ &  \multirow{2}{*}{5}& $i D_\nu H^\dagger D_\mu H B^{\mu\nu}+ \hc$ & \multirow{2}{*}{$\frac{0.04}{E_\text{TeV}^3}, \frac{0.1}{E_\text{TeV}^4}$} \\
2& $i  h W^{+ \al} W^{- \be} \widetilde{F}_{\alpha\beta} + \hc$ & $-$& & $i  D^\be H^\dagger D^{ \al} H  \widetilde{B}_{\alpha\beta} + \hc$&  \\
\hline
3& $i h D^\mu W^{+}_{\nu}  \overset\leftrightarrow{D}_\ga W^-_\mu F^{\nu\ga}+ \hc$ & $+$ &  \multirow{4}{*}{7}&  $i  D^\mu_{\ \nu} H^\dagger  \overset\leftrightarrow{D}_\ga D_\mu H B^{\nu\ga}+ \hc$ & \multirow{4}{*}{$\frac{2 \cdot 10^{-3}}{E_\text{TeV}^5}, \frac{7\cdot 10^{-3}}{E_\text{TeV}^6}$} \\
4& $h D^\mu W^{+}_{\nu}  \overset\leftrightarrow{D}_\ga W^-_\mu F^{\nu\ga}+ \hc$ & $-$ && $  D^\mu_{\ \nu} H^\dagger  \overset\leftrightarrow{D}_\ga D_\mu H B^{\nu\ga}+ \hc$   & \\
5& $i h D_\mu W^{+ \nu}  \overset\leftrightarrow{D}_\ga W^-_\nu F^{\mu\ga}+ \hc$ & $+$ &&  $i  D^\mu_{\ \nu} H^\dagger  \overset\leftrightarrow{D}_\ga D_\nu H B^{\mu\ga}+ \hc$ &  \\
6& $i h  W^{+}_{\mu}  \overset\leftrightarrow{D}_\ga W^-_\nu \d^\mu F^{\nu\ga}+ \hc$ & $+$& & $  i D_\mu H^\dagger  \overset\leftrightarrow{D}_\ga D_\nu H \partial^\mu B^{\nu\ga}+ \hc$  &  \\
\hline
\end{tabular}
\end{center}
\end{adjustwidth}
\caption{Primary operators for $hWW\ga$.}
\label{tab:hWWgamma}
\end{table}

\begin{table}[H]
\scriptsize
\begin{adjustwidth}{-.5in}{-.5in} 
\begin{center}
\renewcommand{\arraystretch}{1.4}
\tabcolsep8pt\begin{tabular}{|c|c|c|c|c|c|}
\hline
 \multirow{2}{*}{$i$} &  \multirow{2}{*}{${\cal{O}}_i^{h ZZ\ga}$}   & \multirow{2}{*}{CP}& \multirow{2}{*}{$d_{\scr{O}_i}$}  & SMEFT & $c$ Unitarity \\[-5pt]
     &    & & & Operator & Bound\\
\hline
1& $ h Z_\mu \overset\leftrightarrow{\d}_\ga Z_\nu \d^\ga F^{\mu\nu}$ & $-$  & \multirow{3}{*}{7}&$ D_\mu H^\dagger  \overset\leftrightarrow{D}_\ga D_\nu H \d^\ga B^{\mu\nu}$  & \multirow{3}{*}{$\frac{3 \cdot 10^{-3}}{E_\text{TeV}^5}, \frac{0.01}{E_\text{TeV}^6}$} \\
2& $ h Z^{\al} \overset\leftrightarrow{\d}_\rho Z^{\be} \d^\rho \widetilde{F}_{\alpha\beta} $ & $+$ && $  D^{\al} H^\dagger \overset\leftrightarrow{D}_\rho D^\be H\d^\rho \widetilde{B}_{\alpha\beta} +\hc $& \\
3& $h \d^\mu Z_{\nu}  \overset\leftrightarrow{\d}_\ga Z_\mu F^{\nu\ga} $ & $-$& & $D^\mu_{\ \nu} H^\dagger \overset\leftrightarrow{D}_\ga D_{\mu} H  B^{\nu\ga}  + \hc$ &  \\
\hline
4& $h  \d^\nu Z_{\mu \ga}  Z_\nu F^{\ga\mu} $ & $-$ & \multirow{2}{*}{7}&  $ iD^\nu W^a_{\mu \ga} H^\dagger \sigma^a  D_\nu H B^{\ga\mu} + h.c.$  &  \multirow{2}{*}{$\frac{0.03}{E_\text{TeV}^4}, \frac{0.3}{E_\text{TeV}^6}$}\\
5& $h  Z_{\ga\mu}  Z_\nu \d^\nu F^{\mu\ga} $ & $-$ &&  $i W^a_{\ga\mu}  H^\dagger \sigma^a D_\nu H \d^\nu B^{\mu\ga} + h.c.$ &  \\
\hline 
6& $ h \widetilde{Z}_{ \gamma \delta}\overset\leftrightarrow{\d}_\rho Z^{\ga} F^{\de\rho} $ & $+$&  \multirow{3}{*}{7} & $i  \widetilde{W}^{a}_{\gamma\delta}  \overset\leftrightarrow{D}_\rho (H^\dagger \si^a D^{\ga} H )B^{\de\rho} +\hc$ & \multirow{3}{*}{$\frac{0.03}{E_\text{TeV}^4}, \frac{0.3}{E_\text{TeV}^6}$} \\
\hline
\end{tabular}
\end{center}
\end{adjustwidth} 
\caption{Primary operators for $hZZ\ga$.}
\label{tab:hZZgamma}
\end{table}

\begin{table}[H]
\scriptsize
\begin{center}
\renewcommand{\arraystretch}{1.4}
\tabcolsep8pt\begin{tabular}{|c|c|c|c|c|c|c|}
\hline
 \multirow{2}{*}{$i$} &  \multirow{2}{*}{${\cal{O}}_i^{h WWZ}$}   & \multirow{2}{*}{CP} & \multirow{2}{*}{$d_{\scr{O}_i}$} & SMEFT & $c$ Unitarity \\[-5pt]
     &    & & & Operator & Bound\\
     \hline
      1 & $i h D^{\nu} W^{+ \mu} W^-_{\mu} Z_{\nu}+\hc$   & $+$& \multirow{5}{*}{5} &  $ i D^{\mu} H^\dagger \si^a D^{\nu}H W^{a}_{\mu\nu}$&  \multirow{5}{*}{$\frac{0.04}{E_\text{TeV}^{3}}, \frac{1}{E_\text{TeV}^6}$} \\
 2 & $i h D^{\mu} W^{+ \nu} W^-_{\mu} Z_{\nu}+\hc$   & $+$ & &  $ i D^{\mu} H^\dagger \si^a D^{\nu}H W^{a}_{\mu\nu}$& \\
   3 & $i h Z^{\mu\nu} W^+_{\mu} W^-_{\nu}$  & $+$& & $ i  D_\mu H^\dagger \si^a D_\nu H  W^{a\mu\nu} $&   \\
4 &$i h \widetilde{W}^+_{\mu\nu} W^{-\mu} Z^\nu + \hc$
& $-$ & & $i  D^{\mu} H^\dagger \si^a D^{\nu}H \widetilde{W}^{a}_{\mu\nu}  $ &    \\
5 & $i h \widetilde{Z}_{\mu\nu} W^{+\mu} W^{-\nu} + \hc$
&   $-$ &&$i  D^{\mu} H^\dagger \si^a D^{\nu}H \widetilde{W}^{a}_{\mu\nu}  $  &    \\
 \hline
6 &$h \widetilde{W}^+_{\mu\nu} W^{-\mu} Z^\nu + \hc$
&   $+$ &  5&  $H^\dagger \si^a   \overset\leftrightarrow{D}{}^{\mu} H H^\dagger  \overset\leftrightarrow{D}{}^{\nu} H  \widetilde{W}^{a}_{\mu\nu}  $
 & $\frac{0.04}{E_\text{TeV}^{3}}, \frac{1}{E_\text{TeV}^6}$
\\
\hline
\end{tabular}
\begin{minipage}{5.7in}
\medskip
\caption{Primary operators for $hWWZ$.  
\label{tab:hWWZ}}
\end{minipage}
\end{center}
\end{table}

\begin{table}[H]
\scriptsize
\begin{center}
\renewcommand{\arraystretch}{1.4}
\tabcolsep5pt\begin{tabular}{|c|c|c|c|c|c|c|}
\hline
 \multirow{2}{*}{$i$} &  \multirow{2}{*}{${\cal{O}}_i^{hZZZ}$}   & \multirow{2}{*}{CP} & \multirow{2}{*}{$d_{\scr{O}_i}$} & SMEFT & $c$ Unitarity \\[-5pt]
     &    & && Operator & Bound\\
\hline
 1 & $h \d^{\mu} Z^{\nu} Z_{\mu} Z_{\nu}$ &   $-$ &5&  $i H^\dagger D^{\mu \nu} H H^\dagger D_{\mu \nu} H  + h.c.$ &  $\frac{5\cdot 10^{-3}}{E_\text{TeV}^4}, \frac{0.04}{E_\text{TeV}^6}$\\[4pt]
 \hline
2 & $ h \d^\mu  Z^{\al\be} \widetilde{Z}_{\alpha\beta}Z_{\mu}$ &   $+$&7 &  $  i D^\mu W^{a\al\be} \widetilde{W}^{a}_{\alpha\beta} H^\dagger D_{\mu} H + h.c.$   & $\frac{0.03}{E_\text{TeV}^4}, \frac{0.3}{E_\text{TeV}^6}$  \\[4pt]
\hline
3 & $h (\widetilde{Z}_{\alpha\beta}  \overset\leftrightarrow{\d}_\mu  Z^{\alpha}) \d^\mu Z^{\beta}$ &   $+$ & 7& $ \widetilde{W}^a_{\alpha\beta} \overset\leftrightarrow{D}_\mu  (D^{\alpha} H^\dagger \sigma^a D^{\mu \beta} H) + h.c.$& $\frac{3 \cdot 10^{-3}}{E_\text{TeV}^5}, \frac{0.01}{E_\text{TeV}^6}$  \\
\hline
4 & $h \d^{\mu \nu} Z^{\rho} \d_\rho Z_{\mu} Z_{\nu}$ &   $-$ &\multirow{4}{*}{7}& $i H^\dagger  D^{\mu \nu \rho} H H^\dagger D_{\rho \mu \nu} H + h.c. $& \multirow{4}{*}{$\frac{3 \cdot 10^{-4}}{E_\text{TeV}^6}, \frac{3\cdot 10^{-3}}{E_\text{TeV}^8}$} \\
5 & $h \d^{\mu} Z^{\nu} \d_\nu Z^{\rho} \d_\rho Z_{\mu}$ &   $-$ && $i H^\dagger  D^{\mu \nu} H  D_\nu^{\ \rho}  H^\dagger  D_{\rho \mu} H + h.c. $ &  \\
6 & $h (\d^{\mu} Z^{\nu}  \overset\leftrightarrow{\d}_\rho  Z_{\mu}) \d^\rho Z_{\nu}$  & $-$& & $ i(H^\dagger D^{\mu \nu} H)  \overset\leftrightarrow{\d}_\rho ( D_{\mu}  H^\dagger D^\rho_{\ \nu} H) + h.c. $  &  \\
\hline
\end{tabular}
\caption{Primary operators for $hZZZ$.
\label{tab:hZZZ}}
\end{center}
\end{table}

\begin{table}[H]
\footnotesize
\begin{center}
\renewcommand{\arraystretch}{1.4}
\tabcolsep10pt\begin{tabular}{|c|c|c|c|c|c|c|}
\hline
 \multirow{2}{*}{$i$} &  \multirow{2}{*}{${\cal{O}}_i^{hh\bar{f}f}$}   & \multirow{2}{*}{CP}& \multirow{2}{*}{$d_{\scr{O}_i}$}  & SMEFT & $c$ Unitarity \\[-5pt]
     &    & & & Operator & Bound\\
\hline
 1 & $h^2 \bar{\psi}_L \psi_R + \hc $ & + & \multirow{2}{*}{5} & $H^\dagger H \bar{Q}_L \tilde H u_R +\hc$ & \multirow{2}{*}{ $\frac{6}{E_\text{TeV}}, \frac{200}{E_\text{TeV}^4}$}\\
 2 & $i h^2 \bar{\psi}_L   \psi_R+ \hc$ & $-$& & $iH^\dagger H \bar{Q}_L \tilde H u_R +\hc$ & \\
 \hline
 3 & $i \d^\mu h\d^\nu h  \bar{\psi}_L\ga_\mu \overset{\leftrightarrow}{\d}_\nu \psi_L $ & + & \multirow{2}{*}{8} & $i D^\mu H^\dagger D^\nu H  \bar{Q}_L\ga_\mu \overset\leftrightarrow{D}_\nu  Q_L$&  \multirow{2}{*}{ $\frac{0.09}{E_\text{TeV}^4}$}  \\
 4 & $i \d^\mu h\d^\nu h  \bar{\psi}_R\ga_\mu \overset{\leftrightarrow}{\d}_\nu \psi_R $ &  + & &  $i D^\mu H^\dagger D^\nu H  \bar{u}_R\ga_\mu \overset\leftrightarrow{D}_\nu  u_R$ & \\
 \hline
\end{tabular}
\caption{Primary operators for $hh\bar{f}f$.
\label{tab:hhff}}
\end{center}
\end{table}

\begin{table}[H]
\scriptsize
\begin{center}
\renewcommand{\arraystretch}{1.4}
\tabcolsep10pt\begin{tabular}{|c|c|c|c|c|c|c|}
\hline
 \multirow{2}{*}{$i$} &  \multirow{2}{*}{${\cal{O}}_i^{hhWW}$}   & \multirow{2}{*}{CP} & \multirow{2}{*}{$d_{\scr{O}_i}$} & SMEFT & $c$ Unitarity \\[-5pt]
     &    & & &Operator & Bound\\
\hline
 1 & $h^2 W^+_\mu W^{- \mu} $ & + & 4&  $H^\dagger H D_\mu H^\dagger D^\mu H$&$ \frac{0.2}{E_\text{TeV}^2}, \frac{2}{E_\text{TeV}^4}$ \\[4pt]
\hline
 2 & $h^2 W^+_{\mu\nu} W^{- \mu\nu}$ & + & \multirow{2}{*}{6}& $H^\dagger H W_{\mu\nu}^a W^{a\mu\nu}$ &  \multirow{2}{*}{$\frac{2}{E_\text{TeV}^2}, \frac{10}{E_\text{TeV}^4}$}\\
 3 & $h^2   W^+_{\mu\nu} \widetilde{W}^{- \mu\nu}$ &  $-$ &&  $H^\dagger H W_{\mu\nu}^a \widetilde{W}^{a\mu\nu}$ & \\
 \hline
 4 & $h  \d^{\mu\nu} h W^+_\mu W^-_{\nu}$ &  + &6& $H^\dagger D^{\mu\nu} H D_\mu H^\dagger D_\nu H + \hc$ &   $\frac{0.01}{E_\text{TeV}^4}, \frac{0.09}{E_\text{TeV}^6}$\\[4pt]
\hline
 5 & $i h \d^{\mu\nu} h \widetilde{W}^+_{\mu\be} W^{-\; \be}_{\nu}+ \hc$ &  $+$& 8& $\ep^{abc}H^\dagger \si^a D^{\mu\nu} H \widetilde{W}^b_{\mu \be} W_\nu^{c\be} + \hc$ & $\frac{0.09}{E_\text{TeV}^4}, \frac{0.9}{E_\text{TeV}^6}$ \\[4pt]
 \hline
 6 & $h \d^{\mu\nu} h  D_\mu W^+_{\be} W_{\nu}^{-\; \be} + \hc$ &  + &  \multirow{4}{*}{8}&  $H^\dagger D^{\mu\nu} H H^\dagger \si^a D_{\mu\be} H  W_\nu^{a\be} + \hc$ &   \multirow{4}{*}{$\frac{7\cdot 10^{-3}}{E_\text{TeV}^5}, \frac{0.2}{E_\text{TeV}^8}$}   \\
\hline 
\end{tabular}
\begin{minipage}{5.7in}
\medskip
\caption{Primary operators for $hhWW$, $hhZ\ga$, and $hh\ga\ga$.
As noted in the text,
the $hh ZZ$ primary operators can be obtained from 
the $hh WW$ operators by the replacement $W_\mu \to Z_\mu$ and the $hh gg$ primary operators can be obtained from 
the $hh \ga\ga$ operators by the replacement $F_{\mu\nu} \to G_{\mu\nu}$. 
\label{tab:hhVV}}
\end{minipage}
\end{center}
\end{table}

\begin{table}[H]
\scriptsize
\begin{center}
\renewcommand{\arraystretch}{1.4}
\tabcolsep8pt\begin{tabular}{|c|c|c|c|c|c|}
\hline
  \multirow{2}{*}{$i$} &  \multirow{2}{*}{${\cal{O}}_i^{hhhZ}$}   & \multirow{2}{*}{CP}& \multirow{2}{*}{$d_{\scr{O}_i}$}  & SMEFT & $c$ Unitarity \\[-5pt]
     &    & & &Operator & Bound\\
\hline
1& $ \left(h  \overset\leftrightarrow{\d}_\nu  \overset\leftrightarrow{\d}_\mu  h \right) \d^\nu h Z^\mu$ & $-$ &7 & $ i H^
\dagger \overset\leftrightarrow{D}_\nu  \overset\leftrightarrow{D}_\mu  H H^\dagger D^{\mu\nu} H + \hc$ & $\frac{0.03}{E_\text{TeV}^4}, \frac{0.3}{E_\text{TeV}^6}$ \\[4pt]
\hline 
2& $ \left(\d^\ga h  \d_\ga \overset\leftrightarrow{\d}_\nu  \overset\leftrightarrow{\d}_\mu  h \right) \d^\nu h Z^\mu$ & $-$&  9& $ i D^\ga H^
\dagger  D_\ga\overset\leftrightarrow{D}_\nu  \overset\leftrightarrow{D}_\mu  H H^\dagger D^{\mu\nu} H + \hc$ & $\frac{2 \cdot 10^{-3}}{E_\text{TeV}^6}, \frac{0.02}{E_\text{TeV}^8}$ \\[4pt]
\hline
3& $ \ep_{\al\be\ga\de} \d^{\eta\si\al} h \d_{\eta\si} \overset\leftrightarrow{\d}_\rho \d^\be h \d^\rho \d^\ga h Z^\de $ & $+$ &13 & $ \ep_{\al\be\ga\de} D^{\eta\si\al} H^\dagger  D_{\eta\si} \overset\leftrightarrow{D}_\rho  D^\be H \ D^\de H^\dagger D^{\rho\ga} H  + \hc$ & $\frac{8\cdot 10^{-6}}{E_\text{TeV}^{10}}$ \\[4pt]
\hline \hline
 \multirow{2}{*}{$i$} &  \multirow{2}{*}{${\cal{O}}_i^{hhh\ga}$}   & \multirow{2}{*}{CP} & \multirow{2}{*}{$d_{\scr{O}_i}$}  & SMEFT & $c$ Unitarity \\[-5pt]
     &    & & & Operator & Bound\\
\hline
1& $ \d^{\eta\si\al} h \d_{\eta\si} \overset\leftrightarrow{\d}_\rho \d^\be h \d^\rho h \widetilde{F}_{\alpha\beta} $ & $+$&   \multirow{2}{*}{13}& $ D^{\eta\si\al} H^\dagger  D_{\eta\si} \overset\leftrightarrow{D}_\rho D^\be H H^\dagger D^\rho H \widetilde{B}_{\alpha\beta} + \hc$& \multirow{2}{*}{$\frac{8 \cdot 10^{-5}}{E_\text{TeV}^9}, \frac{8\cdot 10^{-4}}{E_\text{TeV}^{11}}$} \\
\hline 
\end{tabular}
\caption{Primary operators for $hhhZ$ and $hhh\ga$.
\label{tab:hhhV}}
\end{center}
\end{table}

\section{Conclusions}
\label{sec:conclusions}
This paper has analyzed the most general observables that parameterize
the indirect effect of new heavy physics at colliders.
An important conceptual point is that the space of these observables is
finite, with a finite basis that can be enumerated.
This can be most easily seen in the language of on-shell amplitudes:
any local amplitude can be written as a linear combination of a finitely many
`primary' amplitudes, each of which is multiplied by an infinite
series in Mandelstam invariants.
Under very general physical assumptions, the additional Mandelstam
invariants are suppressed by powers of a heavy mass scale $M$,
and the leading approximation is given by the first nonzero term
in this expansion.
Each primary amplitude can be associated with a local operator,
up to the usual ambiguities from integration by parts and integration
by parts. 
However, these ambiguities do not change the on-shell amplitude, so
we can make the simplest choice when defining the operator basis.

The major results of this paper are a systematic method for determining all
primary operators, and an explicit determination of the 3-point and 4-point
primary operators relevant for Higgs signals at colliders.
The 3-point on-shell amplitudes have no Mandelstam invariants, so there is a
finite list of 3-point operators,
which has previously been found in the literature \Refs{Shadmi:2018xan,Durieux:2019eor,Helset:2020yio}.
Partial results for primary 4-point functions have been given in \cite{Durieux:2019eor, Durieux:2020gip},
and our results agree where they overlap.

The correspondence between local on-shell amplitudes and EFT operators has been
invaluable in this work.
For example, we found that if the on-shell amplitudes are expressed in a 
specific set of kinematic variables, the amplitudes can be treated as
polynomials in the kinematic variables for purposes of determining the linearly
independent amplitudes.
This allows us to efficiently and reliably determine the independent amplitudes.
The Hilbert series that counts independent operators 
is also an invaluable check on these methods.

The primary operators are a natural set of observables for searches for new
physics at colliders, and they can be matched onto theoretical models or EFT
frameworks  (such as SMEFT or HEFT).
We have considered the unitarity and precision electroweak constraints on
these observables, and made a first pass at determining which may be
promising for searches for new physics in Higgs decays.
In particular, the three-body decays into $Z\bar{f}f$, $W\bar{f}f'$, $\ga \bar{f}f$, 
and $Z\ga\ga$ are estimated to be of interest at the HL-LHC.
Investigating the phenomenology of these observables is an obvious
direction for future work.

It is our hope that this framework will prove useful for the LHC program
of constraining (or discovering!) the indirect effects of new 
particles too heavy to be produced.
Under the general assumptions made here, the primary observables are
independent of each other, and experiments can measure them without worrying
about correlations with other observables.
These results can then be compared with predictive theoretical frameworks.
In subsequent work, we plan to study experimental strategies for
carrying out such searches and reporting the results in a way that can be
compared with searches in other channels, or in future colliders.

\textbf{Note:} As this paper was being finalized,  
Ref.~\cite{Dong:2022jru} appeared on the arXiv,
which overlaps with some of our results.
This paper presents a general method for obtaining independent HEFT operators,
but does not distinguish between primaries and descendants.
They give results for operators only up to dimension 8,
and these operators are in agreement with our results.

\section*{Acknowledgements}
We thank
Timothy Cohen,
Nathaniel Craig,
Christophe Grojean,
Xiaochuan Lu,
Peter Onyisi,
Yael Shadmi,
and Zhengkang Zhang
for helpful discussions.
The work of SC was supported by DOE Grant Number DE-SC0011640,
and the work of MC, DL, and MAL was supported by DOE
Grant Number DE-SC-0009999.
The work of SC and MAL was partially performed at the Aspen Center for Physics, 
which is supported by National Science Foundation grant PHY-1607611.

\clearpage

\bibliographystyle{utphys}
\bibliography{references}

\providecommand{\href}[2]{#2}\begingroup\raggedright\begin{thebibliography}{10}

\bibitem{Durieux:2019eor}
G.~Durieux, T.~Kitahara, Y.~Shadmi, and Y.~Weiss, ``{The electroweak effective
  field theory from on-shell amplitudes},''
  \href{http://dx.doi.org/10.1007/JHEP01(2020)119}{{\em JHEP} {\bfseries 01}
  (2020) 119}, \href{http://arxiv.org/abs/1909.10551}{{\ttfamily
  arXiv:1909.10551 [hep-ph]}}.

\bibitem{Durieux:2020gip}
G.~Durieux, T.~Kitahara, C.~S. Machado, Y.~Shadmi, and Y.~Weiss,
  ``{Constructing massive on-shell contact terms},''
  \href{http://dx.doi.org/10.1007/JHEP12(2020)175}{{\em JHEP} {\bfseries 12}
  (2020) 175}, \href{http://arxiv.org/abs/2008.09652}{{\ttfamily
  arXiv:2008.09652 [hep-ph]}}.

\bibitem{Elvang:2010jv}
H.~Elvang, D.~Z. Freedman, and M.~Kiermaier, ``{A simple approach to
  counterterms in N=8 supergravity},''
  \href{http://dx.doi.org/10.1007/JHEP11(2010)016}{{\em JHEP} {\bfseries 11}
  (2010) 016}, \href{http://arxiv.org/abs/1003.5018}{{\ttfamily arXiv:1003.5018
  [hep-th]}}.

\bibitem{Shadmi:2018xan}
Y.~Shadmi and Y.~Weiss, ``{Effective Field Theory Amplitudes the On-Shell Way:
  Scalar and Vector Couplings to Gluons},''
  \href{http://dx.doi.org/10.1007/JHEP02(2019)165}{{\em JHEP} {\bfseries 02}
  (2019) 165}, \href{http://arxiv.org/abs/1809.09644}{{\ttfamily
  arXiv:1809.09644 [hep-ph]}}.

\bibitem{Ma:2019gtx}
T.~Ma, J.~Shu, and M.-L. Xiao, ``{Standard Model Effective Field Theory from
  On-shell Amplitudes},'' \href{http://arxiv.org/abs/1902.06752}{{\ttfamily
  arXiv:1902.06752 [hep-ph]}}.

\bibitem{Durieux:2019siw}
G.~Durieux and C.~S. Machado, ``{Enumerating higher-dimensional operators with
  on-shell amplitudes},''
  \href{http://dx.doi.org/10.1103/PhysRevD.101.095021}{{\em Phys. Rev. D}
  {\bfseries 101} no.~9, (2020) 095021},
  \href{http://arxiv.org/abs/1912.08827}{{\ttfamily arXiv:1912.08827
  [hep-ph]}}.

\bibitem{AccettulliHuber:2021uoa}
M.~Accettulli~Huber and S.~De~Angelis, ``{Standard Model EFTs via on-shell
  methods},'' \href{http://dx.doi.org/10.1007/JHEP11(2021)221}{{\em JHEP}
  {\bfseries 11} (2021) 221}, \href{http://arxiv.org/abs/2108.03669}{{\ttfamily
  arXiv:2108.03669 [hep-th]}}.

\bibitem{Giudice:2007fh}
G.~F. Giudice, C.~Grojean, A.~Pomarol, and R.~Rattazzi, ``{The
  Strongly-Interacting Light Higgs},''
  \href{http://dx.doi.org/10.1088/1126-6708/2007/06/045}{{\em JHEP} {\bfseries
  06} (2007) 045}, \href{http://arxiv.org/abs/hep-ph/0703164}{{\ttfamily
  arXiv:hep-ph/0703164}}.

\bibitem{Azatov:2015oxa}
A.~Azatov, R.~Contino, G.~Panico, and M.~Son, ``{Effective field theory
  analysis of double Higgs boson production via gluon fusion},''
  \href{http://dx.doi.org/10.1103/PhysRevD.92.035001}{{\em Phys. Rev. D}
  {\bfseries 92} no.~3, (2015) 035001},
  \href{http://arxiv.org/abs/1502.00539}{{\ttfamily arXiv:1502.00539
  [hep-ph]}}.

\bibitem{Liu:2016idz}
D.~Liu, A.~Pomarol, R.~Rattazzi, and F.~Riva, ``{Patterns of Strong Coupling
  for LHC Searches},'' \href{http://dx.doi.org/10.1007/JHEP11(2016)141}{{\em
  JHEP} {\bfseries 11} (2016) 141},
  \href{http://arxiv.org/abs/1603.03064}{{\ttfamily arXiv:1603.03064
  [hep-ph]}}.

\bibitem{Buchmuller:1985jz}
W.~Buchmuller and D.~Wyler, ``{Effective Lagrangian Analysis of New
  Interactions and Flavor Conservation},''
  \href{http://dx.doi.org/10.1016/0550-3213(86)90262-2}{{\em Nucl. Phys. B}
  {\bfseries 268} (1986) 621--653}.

\bibitem{Grzadkowski:2010es}
B.~Grzadkowski, M.~Iskrzynski, M.~Misiak, and J.~Rosiek, ``{Dimension-Six Terms
  in the Standard Model Lagrangian},''
  \href{http://dx.doi.org/10.1007/JHEP10(2010)085}{{\em JHEP} {\bfseries 10}
  (2010) 085}, \href{http://arxiv.org/abs/1008.4884}{{\ttfamily arXiv:1008.4884
  [hep-ph]}}.

\bibitem{Brivio:2017vri}
I.~Brivio and M.~Trott, ``{The Standard Model as an Effective Field Theory},''
  \href{http://dx.doi.org/10.1016/j.physrep.2018.11.002}{{\em Phys. Rept.}
  {\bfseries 793} (2019) 1--98},
  \href{http://arxiv.org/abs/1706.08945}{{\ttfamily arXiv:1706.08945
  [hep-ph]}}.

\bibitem{Contino:2013kra}
R.~Contino, M.~Ghezzi, C.~Grojean, M.~Muhlleitner, and M.~Spira, ``{Effective
  Lagrangian for a light Higgs-like scalar},''
  \href{http://dx.doi.org/10.1007/JHEP07(2013)035}{{\em JHEP} {\bfseries 07}
  (2013) 035}, \href{http://arxiv.org/abs/1303.3876}{{\ttfamily arXiv:1303.3876
  [hep-ph]}}.

\bibitem{Elias-Miro:2013mua}
J.~Elias-Miro, J.~R. Espinosa, E.~Masso, and A.~Pomarol, ``{Higgs windows to
  new physics through d=6 operators: constraints and one-loop anomalous
  dimensions},'' \href{http://dx.doi.org/10.1007/JHEP11(2013)066}{{\em JHEP}
  {\bfseries 11} (2013) 066}, \href{http://arxiv.org/abs/1308.1879}{{\ttfamily
  arXiv:1308.1879 [hep-ph]}}.

\bibitem{Alonso:2013hga}
R.~Alonso, E.~E. Jenkins, A.~V. Manohar, and M.~Trott, ``{Renormalization Group
  Evolution of the Standard Model Dimension Six Operators III: Gauge Coupling
  Dependence and Phenomenology},''
  \href{http://dx.doi.org/10.1007/JHEP04(2014)159}{{\em JHEP} {\bfseries 04}
  (2014) 159}, \href{http://arxiv.org/abs/1312.2014}{{\ttfamily arXiv:1312.2014
  [hep-ph]}}.

\bibitem{Henning:2014wua}
B.~Henning, X.~Lu, and H.~Murayama, ``{How to use the Standard Model effective
  field theory},'' \href{http://dx.doi.org/10.1007/JHEP01(2016)023}{{\em JHEP}
  {\bfseries 01} (2016) 023}, \href{http://arxiv.org/abs/1412.1837}{{\ttfamily
  arXiv:1412.1837 [hep-ph]}}.

\bibitem{Feruglio:1992wf}
F.~Feruglio, ``{The Chiral approach to the electroweak interactions},''
  \href{http://dx.doi.org/10.1142/S0217751X93001946}{{\em Int. J. Mod. Phys. A}
  {\bfseries 8} (1993) 4937--4972},
  \href{http://arxiv.org/abs/hep-ph/9301281}{{\ttfamily arXiv:hep-ph/9301281}}.

\bibitem{Alonso:2012px}
R.~Alonso, M.~B. Gavela, L.~Merlo, S.~Rigolin, and J.~Yepes, ``{The Effective
  Chiral Lagrangian for a Light Dynamical ''Higgs Particle''},''
  \href{http://dx.doi.org/10.1016/j.physletb.2013.04.037}{{\em Phys. Lett. B}
  {\bfseries 722} (2013) 330--335},
  \href{http://arxiv.org/abs/1212.3305}{{\ttfamily arXiv:1212.3305 [hep-ph]}}.
  [Erratum: Phys.Lett.B 726, 926 (2013)].

\bibitem{Buchalla:2013rka}
G.~Buchalla, O.~Cat\`a, and C.~Krause, ``{Complete Electroweak Chiral
  Lagrangian with a Light Higgs at NLO},''
  \href{http://dx.doi.org/10.1016/j.nuclphysb.2014.01.018}{{\em Nucl. Phys. B}
  {\bfseries 880} (2014) 552--573},
  \href{http://arxiv.org/abs/1307.5017}{{\ttfamily arXiv:1307.5017 [hep-ph]}}.
  [Erratum: Nucl.Phys.B 913, 475--478 (2016)].

\bibitem{Brivio:2016fzo}
I.~Brivio, J.~Gonzalez-Fraile, M.~C. Gonzalez-Garcia, and L.~Merlo, ``{The
  complete HEFT Lagrangian after the LHC Run I},''
  \href{http://dx.doi.org/10.1140/epjc/s10052-016-4211-9}{{\em Eur. Phys. J. C}
  {\bfseries 76} no.~7, (2016) 416},
  \href{http://arxiv.org/abs/1604.06801}{{\ttfamily arXiv:1604.06801
  [hep-ph]}}.

\bibitem{Sun:2022ssa}
H.~Sun, M.-L. Xiao, and J.-H. Yu, ``{Complete NLO Operators in the Higgs
  Effective Field Theory},'' \href{http://arxiv.org/abs/2206.07722}{{\ttfamily
  arXiv:2206.07722 [hep-ph]}}.

\bibitem{Gupta:2014rxa}
R.~S. Gupta, A.~Pomarol, and F.~Riva, ``{BSM Primary Effects},''
  \href{http://dx.doi.org/10.1103/PhysRevD.91.035001}{{\em Phys. Rev. D}
  {\bfseries 91} no.~3, (2015) 035001},
  \href{http://arxiv.org/abs/1405.0181}{{\ttfamily arXiv:1405.0181 [hep-ph]}}.

\bibitem{Falkowski:2001958}
A.~Falkowski, ``{Higgs Basis: Proposal for an EFT basis choice for LHC
  HXSWG},''. \url{https://cds.cern.ch/record/2001958}.

\bibitem{Gonzalez-Alonso:2014eva}
M.~Gonzalez-Alonso, A.~Greljo, G.~Isidori, and D.~Marzocca,
  ``{Pseudo-observables in Higgs decays},''
  \href{http://dx.doi.org/10.1140/epjc/s10052-015-3345-5}{{\em Eur. Phys. J. C}
  {\bfseries 75} (2015) 128}, \href{http://arxiv.org/abs/1412.6038}{{\ttfamily
  arXiv:1412.6038 [hep-ph]}}.

\bibitem{Gonzalez-Alonso:2015bha}
M.~Gonzalez-Alonso, A.~Greljo, G.~Isidori, and D.~Marzocca, ``{Electroweak
  bounds on Higgs pseudo-observables and $h \to 4 \ell$ decays},''
  \href{http://dx.doi.org/10.1140/epjc/s10052-015-3555-x}{{\em Eur. Phys. J. C}
  {\bfseries 75} (2015) 341}, \href{http://arxiv.org/abs/1504.04018}{{\ttfamily
  arXiv:1504.04018 [hep-ph]}}.

\bibitem{Galloway:2013dma}
J.~Galloway, M.~A. Luty, Y.~Tsai, and Y.~Zhao, ``{Induced Electroweak Symmetry
  Breaking and Supersymmetric Naturalness},''
  \href{http://dx.doi.org/10.1103/PhysRevD.89.075003}{{\em Phys. Rev. D}
  {\bfseries 89} no.~7, (2014) 075003},
  \href{http://arxiv.org/abs/1306.6354}{{\ttfamily arXiv:1306.6354 [hep-ph]}}.

\bibitem{Cohen:2020xca}
T.~Cohen, N.~Craig, X.~Lu, and D.~Sutherland, ``{Is SMEFT Enough?},''
  \href{http://dx.doi.org/10.1007/JHEP03(2021)237}{{\em JHEP} {\bfseries 03}
  (2021) 237}, \href{http://arxiv.org/abs/2008.08597}{{\ttfamily
  arXiv:2008.08597 [hep-ph]}}.

\bibitem{Banta:2021dek}
I.~Banta, T.~Cohen, N.~Craig, X.~Lu, and D.~Sutherland, ``{Non-decoupling new
  particles},'' \href{http://dx.doi.org/10.1007/JHEP02(2022)029}{{\em JHEP}
  {\bfseries 02} (2022) 029}, \href{http://arxiv.org/abs/2110.02967}{{\ttfamily
  arXiv:2110.02967 [hep-ph]}}.

\bibitem{Agrawal:2019bpm}
P.~Agrawal, D.~Saha, L.-X. Xu, J.-H. Yu, and C.~P. Yuan, ``{Determining the
  shape of the Higgs potential at future colliders},''
  \href{http://dx.doi.org/10.1103/PhysRevD.101.075023}{{\em Phys. Rev. D}
  {\bfseries 101} no.~7, (2020) 075023},
  \href{http://arxiv.org/abs/1907.02078}{{\ttfamily arXiv:1907.02078
  [hep-ph]}}.

\bibitem{Chang:2019vez}
S.~Chang and M.~A. Luty, ``{The Higgs Trilinear Coupling and the Scale of New
  Physics},'' \href{http://dx.doi.org/10.1007/JHEP03(2020)140}{{\em JHEP}
  {\bfseries 03} (2020) 140}, \href{http://arxiv.org/abs/1902.05556}{{\ttfamily
  arXiv:1902.05556 [hep-ph]}}.

\bibitem{Falkowski:2019tft}
A.~Falkowski and R.~Rattazzi, ``{Which EFT},''
  \href{http://dx.doi.org/10.1007/JHEP10(2019)255}{{\em JHEP} {\bfseries 10}
  (2019) 255}, \href{http://arxiv.org/abs/1902.05936}{{\ttfamily
  arXiv:1902.05936 [hep-ph]}}.

\bibitem{Abu-Ajamieh:2020yqi}
F.~Abu-Ajamieh, S.~Chang, M.~Chen, and M.~A. Luty, ``{Higgs coupling
  measurements and the scale of new physics},''
  \href{http://dx.doi.org/10.1007/JHEP07(2021)056}{{\em JHEP} {\bfseries 07}
  (2021) 056}, \href{http://arxiv.org/abs/2009.11293}{{\ttfamily
  arXiv:2009.11293 [hep-ph]}}.

\bibitem{Helset:2020yio}
A.~Helset, A.~Martin, and M.~Trott, ``{The Geometric Standard Model Effective
  Field Theory},'' \href{http://dx.doi.org/10.1007/JHEP03(2020)163}{{\em JHEP}
  {\bfseries 03} (2020) 163}, \href{http://arxiv.org/abs/2001.01453}{{\ttfamily
  arXiv:2001.01453 [hep-ph]}}.

\bibitem{Coleman:1969sm}
S.~R. Coleman, J.~Wess, and B.~Zumino, ``{Structure of phenomenological
  Lagrangians. 1.},'' \href{http://dx.doi.org/10.1103/PhysRev.177.2239}{{\em
  Phys. Rev.} {\bfseries 177} (1969) 2239--2247}.

\bibitem{Coleman:1985rnk}
S.~Coleman, \href{http://dx.doi.org/10.1017/CBO9780511565045}{{\em {Aspects of
  Symmetry}: {Selected Erice Lectures}}}.
\newblock Cambridge University Press, Cambridge, U.K., 1985.

\bibitem{Georgi:1991ch}
H.~Georgi, ``{On-shell effective field theory},''
  \href{http://dx.doi.org/10.1016/0550-3213(91)90244-R}{{\em Nucl. Phys. B}
  {\bfseries 361} (1991) 339--350}.

\bibitem{Arzt:1993gz}
C.~Arzt, ``{Reduced effective Lagrangians},''
  \href{http://dx.doi.org/10.1016/0370-2693(94)01419-D}{{\em Phys. Lett. B}
  {\bfseries 342} (1995) 189--195},
  \href{http://arxiv.org/abs/hep-ph/9304230}{{\ttfamily arXiv:hep-ph/9304230}}.

\bibitem{Mathematica}
W.~R. Inc., ``Mathematica, {V}ersion 13.1.''
\newblock \url{https://www.wolfram.com/mathematica}. Champaign, IL, 2022.

\bibitem{Lehman:2015via}
L.~Lehman and A.~Martin, ``{Hilbert Series for Constructing Lagrangians:
  expanding the phenomenologist's toolbox},''
  \href{http://dx.doi.org/10.1103/PhysRevD.91.105014}{{\em Phys. Rev. D}
  {\bfseries 91} (2015) 105014},
  \href{http://arxiv.org/abs/1503.07537}{{\ttfamily arXiv:1503.07537
  [hep-ph]}}.

\bibitem{Henning:2015daa}
B.~Henning, X.~Lu, T.~Melia, and H.~Murayama, ``{Hilbert series and operator
  bases with derivatives in effective field theories},''
  \href{http://dx.doi.org/10.1007/s00220-015-2518-2}{{\em Commun. Math. Phys.}
  {\bfseries 347} no.~2, (2016) 363--388},
  \href{http://arxiv.org/abs/1507.07240}{{\ttfamily arXiv:1507.07240
  [hep-th]}}.

\bibitem{Lehman:2015coa}
L.~Lehman and A.~Martin, ``{Low-derivative operators of the Standard Model
  effective field theory via Hilbert series methods},''
  \href{http://dx.doi.org/10.1007/JHEP02(2016)081}{{\em JHEP} {\bfseries 02}
  (2016) 081}, \href{http://arxiv.org/abs/1510.00372}{{\ttfamily
  arXiv:1510.00372 [hep-ph]}}.

\bibitem{Henning:2015alf}
B.~Henning, X.~Lu, T.~Melia, and H.~Murayama, ``{2, 84, 30, 993, 560, 15456,
  11962, 261485, ...: Higher dimension operators in the SM EFT},''
  \href{http://dx.doi.org/10.1007/JHEP08(2017)016}{{\em JHEP} {\bfseries 08}
  (2017) 016}, \href{http://arxiv.org/abs/1512.03433}{{\ttfamily
  arXiv:1512.03433 [hep-ph]}}. [Erratum: JHEP 09, 019 (2019)].

\bibitem{Henning:2017fpj}
B.~Henning, X.~Lu, T.~Melia, and H.~Murayama, ``{Operator bases, $S$-matrices,
  and their partition functions},''
  \href{http://dx.doi.org/10.1007/JHEP10(2017)199}{{\em JHEP} {\bfseries 10}
  (2017) 199}, \href{http://arxiv.org/abs/1706.08520}{{\ttfamily
  arXiv:1706.08520 [hep-th]}}.

\bibitem{Graf:2020yxt}
L.~Graf, B.~Henning, X.~Lu, T.~Melia, and H.~Murayama, ``{2, 12, 117, 1959,
  45171, 1170086, \textellipsis{}: a Hilbert series for the QCD chiral
  Lagrangian},'' \href{http://dx.doi.org/10.1007/JHEP01(2021)142}{{\em JHEP}
  {\bfseries 01} (2021) 142}, \href{http://arxiv.org/abs/2009.01239}{{\ttfamily
  arXiv:2009.01239 [hep-ph]}}.

\bibitem{Graf:2022rco}
L.~Gr\'af, B.~Henning, X.~Lu, T.~Melia, and H.~Murayama, ``{Hilbert Series, the
  Higgs Mechanism, and HEFT},''
  \href{http://arxiv.org/abs/2211.06275}{{\ttfamily arXiv:2211.06275
  [hep-ph]}}.

\bibitem{Sun:2022aag}
H.~Sun, Y.-N. Wang, and J.-H. Yu, ``{Hilbert Series and Operator Counting on
  the Higgs Effective Field Theory},''
  \href{http://arxiv.org/abs/2211.11598}{{\ttfamily arXiv:2211.11598
  [hep-ph]}}.

\bibitem{DeAngelis:2022qco}
S.~De~Angelis, ``{Amplitude bases in generic EFTs},''
  \href{http://dx.doi.org/10.1007/JHEP08(2022)299}{{\em JHEP} {\bfseries 08}
  (2022) 299}, \href{http://arxiv.org/abs/2202.02681}{{\ttfamily
  arXiv:2202.02681 [hep-th]}}.

\bibitem{Cornwall:1974km}
J.~M. Cornwall, D.~N. Levin, and G.~Tiktopoulos, ``{Derivation of Gauge
  Invariance from High-Energy Unitarity Bounds on the s Matrix},''
  \href{http://dx.doi.org/10.1103/PhysRevD.10.1145}{{\em Phys. Rev. D}
  {\bfseries 10} (1974) 1145}. [Erratum: Phys.Rev.D 11, 972 (1975)].

\bibitem{Liu:2022alx}
D.~Liu and Z.~Yin, ``{Gauge invariance from on-shell massive amplitudes and
  tree-level unitarity},''
  \href{http://dx.doi.org/10.1103/PhysRevD.106.076003}{{\em Phys. Rev. D}
  {\bfseries 106} no.~7, (2022) 076003},
  \href{http://arxiv.org/abs/2204.13119}{{\ttfamily arXiv:2204.13119
  [hep-th]}}.

\bibitem{Abu-Ajamieh:2021egq}
F.~Abu-Ajamieh, ``{The scale of new physics from the Higgs couplings to
  \ensuremath{\gamma}\ensuremath{\gamma} and \ensuremath{\gamma}Z},''
  \href{http://dx.doi.org/10.1007/JHEP06(2022)091}{{\em JHEP} {\bfseries 06}
  (2022) 091}, \href{http://arxiv.org/abs/2112.13529}{{\ttfamily
  arXiv:2112.13529 [hep-ph]}}.

\bibitem{Abu-Ajamieh:2022ppp}
F.~Abu-Ajamieh, ``{The scale of new physics from the Higgs couplings to gg},''
  \href{http://dx.doi.org/10.1016/j.physletb.2022.137389}{{\em Phys. Lett. B}
  {\bfseries 833} (2022) 137389},
  \href{http://arxiv.org/abs/2203.07410}{{\ttfamily arXiv:2203.07410
  [hep-ph]}}.

\bibitem{Vayonakis:1976vz}
C.~E. Vayonakis, ``{Born Helicity Amplitudes and Cross-Sections in Nonabelian
  Gauge Theories},'' \href{http://dx.doi.org/10.1007/BF02746538}{{\em Lett.
  Nuovo Cim.} {\bfseries 17} (1976) 383}.

\bibitem{Chanowitz:1985hj}
M.~S. Chanowitz and M.~K. Gaillard, ``{The TeV Physics of Strongly Interacting
  W's and Z's},'' \href{http://dx.doi.org/10.1016/0550-3213(85)90580-2}{{\em
  Nucl. Phys. B} {\bfseries 261} (1985) 379--431}.

\bibitem{Kribs:2020jgn}
G.~D. Kribs, X.~Lu, A.~Martin, and T.~Tong, ``{Custodial symmetry violation in
  the SMEFT},'' \href{http://dx.doi.org/10.1103/PhysRevD.104.056006}{{\em Phys.
  Rev. D} {\bfseries 104} no.~5, (2021) 056006},
  \href{http://arxiv.org/abs/2009.10725}{{\ttfamily arXiv:2009.10725
  [hep-ph]}}.

\bibitem{Escribano:1993xr}
R.~Escribano and E.~Masso, ``{Constraints on fermion magnetic and electric
  moments from LEP-1},''
  \href{http://dx.doi.org/10.1016/S0550-3213(94)80039-1}{{\em Nucl. Phys. B}
  {\bfseries 429} (1994) 19--32},
  \href{http://arxiv.org/abs/hep-ph/9403304}{{\ttfamily arXiv:hep-ph/9403304}}.

\bibitem{LHCHiggsCrossSectionWorkingGroup:2012nn}
{\bfseries LHC Higgs Cross Section Working Group} Collaboration, A.~David,
  A.~Denner, M.~Duehrssen, M.~Grazzini, C.~Grojean, G.~Passarino,
  M.~Schumacher, M.~Spira, G.~Weiglein, and M.~Zanetti, ``{LHC HXSWG interim
  recommendations to explore the coupling structure of a Higgs-like
  particle},'' \href{http://arxiv.org/abs/1209.0040}{{\ttfamily arXiv:1209.0040
  [hep-ph]}}.

\bibitem{Cepeda:2019klc}
M.~Cepeda {\em et~al.}, ``{Report from Working Group 2}: {Higgs Physics at the
  HL-LHC and HE-LHC},''
  \href{http://dx.doi.org/10.23731/CYRM-2019-007.221}{{\em CERN Yellow Rep.
  Monogr.} {\bfseries 7} (2019) 221--584},
  \href{http://arxiv.org/abs/1902.00134}{{\ttfamily arXiv:1902.00134
  [hep-ph]}}.

\bibitem{Hagiwara:1986vm}
K.~Hagiwara, R.~D. Peccei, D.~Zeppenfeld, and K.~Hikasa, ``{Probing the Weak
  Boson Sector in e+ e- ---\ensuremath{>} W+ W-},''
  \href{http://dx.doi.org/10.1016/0550-3213(87)90685-7}{{\em Nucl. Phys. B}
  {\bfseries 282} (1987) 253--307}.

\bibitem{Boselli:2017pef}
S.~Boselli, C.~M. Carloni~Calame, G.~Montagna, O.~Nicrosini, F.~Piccinini, and
  A.~Shivaji, ``{Higgs decay into four charged leptons in the presence of
  dimension-six operators},''
  \href{http://dx.doi.org/10.1007/JHEP01(2018)096}{{\em JHEP} {\bfseries 01}
  (2018) 096}, \href{http://arxiv.org/abs/1703.06667}{{\ttfamily
  arXiv:1703.06667 [hep-ph]}}.

\bibitem{Dong:2022jru}
Z.-Y. Dong, T.~Ma, J.~Shu, and Z.-Z. Zhou, ``{The New Formulation of Higgs
  Effective Field Theory},'' \href{http://arxiv.org/abs/2211.16515}{{\ttfamily
  arXiv:2211.16515 [hep-ph]}}.

\end{thebibliography}\endgroup

\end{document}